\documentclass[aps,floatfix,amsmath,nofootinbib,amssymb,superscriptaddress,letterpaper,twocolumn]{revtex4}

\usepackage{overpic}
\usepackage{amssymb}
\usepackage{indentfirst}
\usepackage{feynmf}
\usepackage{slashed}
\usepackage{cases}
\usepackage{color}
\usepackage{multirow}
\usepackage{epstopdf}
\usepackage{graphicx,color,bm}
\usepackage{booktabs}
\usepackage{float}
\usepackage{amsmath}
\allowdisplaybreaks[4]

\usepackage[colorlinks=true,
            citecolor=green,
            anchorcolor=red,
            menucolor=red,
            linkcolor=red,
            filecolor=red,
            runcolor=red,
            urlcolor=blue]{hyperref}

\begin{document}

\title{T-odd transverse momentum dependent gluon distributions for tensor polarized deuteron in a spectator model}

\author{Xiupeng Xie}\affiliation{School of Physics, Southeast University, Nanjing
211189, China}

\author{Zhun Lu}
\email{zhunlu@seu.edu.cn}
\affiliation{School of Physics, Southeast University, Nanjing 211189, China}

\begin{abstract}
We present a model calculation of the T-odd transverse momentum dependent distribution functions (TMDs) for gluons inside a tensor polarized deuteron. The model is built on the assumption that an on-shell deuteron can emit a time-like off-shell gluon, with the residual system treated as a single on-shell spectator particle. The spectator mass is described via a spectral function, which allows it to take real values over a continuous range. The final-state interaction required to generate nonvanishing T-odd functions are implemented via single-gluon exchange between the spectator and the outgoing parton. We derive analytical expressions for six T-odd gluon TMDs, and present numerical results characterizing their dependence on the longitudinal momentum fraction $x$ and the transverse momentum $\bm{k}_T$.
\end{abstract}

\maketitle

\section{Introduction}
Elucidating the partonic structure of hadrons is a central goal of modern particle physics.
In high energy collisions, gluons play a increasingly dominant role with rising collision energy, driven by the decreasing typical longitudinal momentum fraction $x$ accessed at higher energies. This kinematic regime has been extensively mapped by inclusive deep inelastic scattering (DIS) experiments at the Hadron-Electron Ring Accelerator (HERA) and proton-proton collision studies at the Large Hadron Collider (LHC). In semi-inclusive and exclusive processes, by contrast, the transverse momentum dependent distribution functions (TMDs) of gluons can be directly constrained. This gives rise to a rich set of gluon TMDs, most notably when accounting for hadron polarization.

For spin-1/2 nucleons, both unpolarized and vector polarized TMDs have been rigorously characterized via decades of complementary experimental measurements and theoretical advances. While the density matrix of a spin-1/2 hadron is fully parametrized by a single spin vector $S$, describing a spin-1 hadron additionally requires a spin tensor $T$. Despite the extensive body of knowledge accumulated on the momentum and coordinate-space distributions of quarks and gluons inside nucleons, the fundamental partonic structure of hadrons with spin greater than 1/2 remains drastically understudied. Research into tensor polarized TMDs is therefore critical to unveiling the internal structure of spin-1 hadrons such as the deuteron. 

Unpolarized and vector polarized gluon TMDs are accessible via dedicated experimental measurements at facilities including the Relativistic Heavy Ion Collider (RHIC), the Large Hadron Collider (LHC), the proposed future polarized fixed-target experiment at the LHC (AFTER@LHC)~\cite{Brodsky:2012vg,Boer:2012bt,Signori:2016jwo,denDunnen:2014kjo}, and an Electron-Ion Collider (EIC)~\cite{Boer:2011fh}. Linearly polarized gluon TMDs can be extracted via measurements of the $cos(2\phi)$ azimuthal modulations in processes including dijet or heavy quark pair production in electro-proton or electron-nucleus collisions~\cite{Pisano:2013cya,Dumitru:2015gaa}, as well as virtual photon-jet pair production in $pp$ or $pA$ collisions~\cite{Metz:2011wb}. Heavy quarkonium production in unpolarized and polarized $pp$ collisions also provides a sensitive channel to access these linearly polarized gluon TMDs, often in coincidence with other gluon TMDs~\cite{Boer:2012bt,Signori:2016jwo}. Finally, Tensor polarized gluon TMDs can be studied via experiments with polarized deuteron beams, including those proposed at the EIC option put forward at Jefferson Lab (JLEIC)~\cite{Boer:2011fh,Abeyratne:2012ah,Abeyratne:2015pma}, or at COMPASS~\cite{Ball:2006zz}, though the accessible small $x$ region remains limited.

In this paper, we perform a systematic investigation of T-odd gluon TMDs in a spectator model. The model is built on the assumption that an on-shell deuteron can emit a time-like off-shell gluon, with the residual system treated as a single on-shell spectator particle. The spectator mass is described via a spectral function, which permits it to take real values over a continuous range. The spectator model was originally developed to describe quark TMDs in nucleons~\cite{Jakob:1997wg,Brodsky:2002cx,Gamberg:2003ey,Bacchetta:2003rz,Bacchetta:2008af,Bacchetta:2010si} and has subsequently been successfully extended in multiple formulations to investigate a broad array of hadronic phenomena. These applications include quark TMDs of the pions~\cite{Lu:2004hu,Meissner:2008ay,Ma:2019agv}, transverse momentum dependent (TMD) fragmentation functions (FFs) of pions and kaons~\cite{Bacchetta:2007wc}, and, more recently, both T-even and T-odd gluon TMDs and TMD FFs~\cite{Lu:2016vqu,Bacchetta:2020vty,Bacchetta:2021lvw,Bacchetta:2021twk,Bacchetta:2024fci,Xie:2022lra,Xie:2024cnv}. Our work extends this well-established theoretical methodology to the case of T-odd gluon TMDs in tensor polarized spin-1 hadronic systems.

In a previous work~\cite{Xie:2026uja}, we computed the complete set of T-even gluon TMDs within a spectator model.
Building on this framework, the present study extends our earlier analysis by performing a systematic calculation of all leading-twist T-odd TMDs in the same model, including a complete characterization of their process dependence.
The spectator-model framework has also been adopted to investigate the Collins function for pions and kaons~\cite{Bacchetta:2007wc,Bacchetta:2001di,Bacchetta:2002tk,Gamberg:2003eg,Bacchetta:2003xn,Amrath:2005gv}, as well as gluon TMD FFs~\cite{Xie:2024cnv}, through incorporating pion loops or gluon loops. 
Moreover, a set of twist-3 fragmentation functions $\tilde{H}\left(z,\bm{k}_T^2\right)$, $H\left(z,\bm{k}_T^2\right)$, $\tilde{G}^\perp \left(z,\bm{k}_T^2\right)$, and $G^\perp\left(z,\bm{k}_T^2\right)$ have been calculated within this model~\cite{Lu:2015wja,Yang:2016mxl}.

The remainder of this paper is organized as follows. In Sec.~\ref{section2}, we outline the spectator-model formalism, including the implementation of gluon rescattering effects. We derive the one-loop gluon-gluon correlator and extract the leading-twist T-odd gluon TMDs using the appropriate projection operators.  Numerical results for these T-odd gluon TMDs are presented in Sec.~\ref{section3},. Finally, we summarize our main results and conclude in Sec.~\ref{section4}.


\section{Analytic calculation of T-odd gluon TMDs}\label{section2}

In a reference frame where the hadron has no transverse momentum, the hadron momentum $P$ and the parton momentum $k$ are given by:
\begin{align}
P^\mu=&P^+ n_+^\mu + \frac{M^2}{2P^+} n_-^\mu \,,\\
k^\mu=&xP^+ n_+^\mu + \frac{k^2+\bm{k}_T^2}{2xP^+} n_-^\mu +\bm{k}_T \,,
\end{align}
where $M$ is the hadron mass and $x=k^+/P^+$ denotes the longitudinal momentum fraction carried by the parton. Here we adopt light-front coordinates, defined as $a^\mu \equiv a^+ n_+^\mu + a^- n_-^\mu +a_T^\mu$, with the transverse component $a^\mu_T$ corresponding to the transverse momentum vector $\bm{a}_T$. The light-cone basis vectors satisfy the standard orthogonality relations $n_+^2=n_-^2=0$ and $n_+ \cdot n_- =1$, with explicit components $n_+\equiv \left[1,0,\bm{0}_T\right]$ and $n_-\equiv \left[0,1,\bm{0}_T\right]$.

Hadron polarization consists of one of the core ingredients in TMD phenomenology. For a spin-1 hadron, the full polarization state is characterized by a spin vector $S^\mu$ and a symmetric traceless spin tensor $T^{\mu \nu}$~\cite{Bacchetta:2000jk}. These satisfy $P \cdot S=0$ and $P_\mu T^{\mu \nu}=0$, and are parametrized in the light-cone basis as
\begin{align}
S^\mu=&S_L \frac{P^+}{M} n_+^\mu-S_L\frac{M}{2P^+} n_-^\mu+ S_T^\mu\,,\\
T^{\mu \nu}=&\frac{1}{2}\Bigg[\frac{4}{3}S_{LL}\frac{\left(P\cdot n_-\right)^2}{M^2} n_+^\mu n_+^\nu-\frac{2}{3}S_{LL}\left(n_+^{\left\{\mu\right.}n_-^{\left.\nu \right\}}-g_T^{\mu \nu}\right)\notag\\
&+\frac{1}{3}S_{LL}\frac{M^2}{\left(P \cdot n_-\right)^2}n_-^\mu n_-^\nu+\frac{P\cdot n_-}{M}n_+^{\left\{\mu\right.} S_{LT}^{\left.\nu\right\}}\notag\\
&-\frac{M}{2P\cdot n_-}n_-^{\left\{\mu\right.} S_{LT}^{\left.\nu\right\}}+S_{TT}^{\mu\nu}\Bigg]\,,
\end{align}
where $g_T^{\mu \nu}=g^{\mu \nu}-n_+^{\left\{\mu\right.}n_-^{\left.\nu \right\}}$ is the symmetric transverse tensor, and the notation $a^{\left\{\mu\right.} b^{\left.\nu \right\}}$ denotes index symmetrization: $a^{\left\{\mu\right.} b^{\left.\nu \right\}}=a^\mu b^\nu +a^\nu b^\mu$. 

The polarization density matrix for a spin-1 hadron admits the standard Lorentz-invariant decomposition:
\begin{align}
&\epsilon^{*\mu}\left(P,\lambda\right) \epsilon^\nu \left(P,\lambda\right)\notag\\
=&-\frac{1}{3}\left(g^{\mu \nu}-\frac{P^\mu P^\nu}{M^2}\right)+\frac{i}{2M}\epsilon^{\mu \nu \alpha \beta}P_\alpha S_\beta-T^{\mu \nu}\,.
\end{align}

The gauge-invariant gluon-gluon correlator for spin-1 hadrons, which encodes all leading-twist gluon TMD information, is defined as~\cite{Mulders:2000sh,Meissner:2007rx}
\begin{align}
\Phi&^{\mu \nu;\rho \sigma}\left(x,\bm{k}_T;S,T\right)=\frac{1}{xP^+}\int \frac{d\xi^- d \bm{\xi}_T}{(2\pi)^3} e^{ik\cdot \xi} \notag\\
&\times \left. \left\langle P,S,T\left|F^{\mu \nu}\left(0\right) U_{[0,\xi]} F^{\rho \sigma}\left(\xi\right) U_{[\xi,0]}^\prime \right|P,S,T\right\rangle\right|_{\xi^+=0}\,,\label{eq:phi0}
\end{align}
where a implicit summation over color indices is performed, and the process-dependent gauge links $U_{[0,\xi]}$ and $U_{[\xi,0]}^\prime$ enforce color gauge invariance of the correlator.

At leading twist, we restrict our attention to the physical component of the gluon–gluon correlator $\Phi^{+i,+j}\equiv \Phi^{ij}$, where $i,j$ run over transverse spatial indices. This correlator permits a full parameterization via the complete set of leading-twist gluon TMDs, as given in Ref.~\cite{Boer:2016xqr}:
\begin{align}
\Phi^{ij}_U \left(x,\bm{k}_T\right)=&\frac{1}{2}\left[ -g_T^{ij} f_1\left(x,\bm{k}_T^2\right)+\frac{k_T^{ij}}{M^2}h_1^\perp \left(x,\bm{k}_T^2\right)\right]\,,\label{eq:phiU}\\
\Phi^{ij}_L \left(x,\bm{k}_T\right)=&\frac{1}{2}\Bigg[ i \epsilon_T^{ij}S_L g_1\left(x,\bm{k}_T^2\right)\notag\\
&+\frac{\epsilon_{T \alpha}^{\left\{i\right.} k_T^{\left.j\right\}\alpha } S_L}{2M^2} h_{1L}^\perp \left(x,\bm{k}_T^2\right)\Bigg]\,,\\
\Phi^{ij}_T \left(x,\bm{k}_T\right)=&\frac{1}{2}\Bigg[-\frac{g_T^{ij}\epsilon_T^{S_T k_T}}{M} f_{1T}^\perp \left(x,\bm{k}_T^2\right)\notag\\
&+\frac{i \epsilon_T^{ij} \bm{k}_T \cdot \bm{S}_T}{M}g_{1T}\left(x,\bm{k}_T^2\right) \notag\\ 
&-\frac{\epsilon_T^{k_T \left\{i\right.} S_T^{\left.j\right\}}+\epsilon_T^{S_T \left\{i\right.} k_T^{\left.j\right\}}}{4M} h_1 \left(x,\bm{k}_T^2\right)\notag\\
&-\frac{\epsilon_{T \alpha}^{\left\{i\right.} k_T^{\left.j\right\}\alpha S_T}}{2M^3} h_{1T}^\perp \left(x,\bm{k}_T^2\right)\Bigg]\,,\\
\Phi^{ij}_{LL} \left(x,\bm{k}_T\right)=&\frac{1}{2}\Bigg[-g_T^{ij} S_{LL} f_{1LL}\left(x,\bm{k}_T^2\right)\notag\\
&+\frac{k_T^{ij} S_{LL}}{M^2}h_{1LL}^\perp \left(x,\bm{k}_T^2\right)\Bigg]\,,\\
\Phi^{ij}_{LT} \left(x,\bm{k}_T\right)=&\frac{1}{2}\Bigg[-\frac{g_T^{ij} \bm{k}_T \cdot \bm{S}_{LT}}{M} f_{1LT}\left(x,\bm{k}_T^2\right)\notag\\
&+\frac{i \epsilon_T^{ij} \epsilon_T^{S_{LT}k_T}}{M}g_{1LT}\left(x,\bm{k}_T^2\right)\notag\\
&+\frac{S_{LT}^{\left\{i\right.} k_T^{\left.j\right\}}}{M} h_{1LT}\left(x,\bm{k}_T^2\right)\notag\\
&+\frac{k_T^{ij\alpha} S_{LT\alpha}}{M^3} h_{1LT}^\perp \left(x,\bm{k}_T^2\right)\Bigg]\,,\\
\Phi^{ij}_{TT} \left(x,\bm{k}_T\right)=&\frac{1}{2}\Bigg[-\frac{g_T^{ij}k_T^{\alpha \beta} S_{TT\alpha \beta}}{M^2} f_{1TT}\left(x,\bm{k}_T^2\right)\notag\\
&+\frac{i \epsilon_T^{ij} \epsilon_{T\gamma}^{\beta} k_T^{\gamma \alpha} S_{TT\alpha \beta}}{M^2} g_{1TT}\left(x,\bm{k}_T^2\right)\notag\\
&+S_{TT}^{ij} h_{1TT}\left(x,\bm{k}_T^2\right)\notag\\
&+\frac{S_{TT \alpha}^{\left\{i\right.} k_T^{\left.j\right\}\alpha}}{M^2}h_{1TT}^\perp \left(x,\bm{k}_T^2\right)\notag\\
&+\frac{k_T^{ij \alpha \beta} S_{TT\alpha \beta}}{M^4} h_{1TT}^{\perp \perp}\left(x,\bm{k}_T^2\right)\Bigg]\,.\label{eq:phiTT}
\end{align}
Here, $\epsilon_T^{ij}=\epsilon^{n_+ n_- ij}=\epsilon^{-+ij}$ is the antisymmetric transverse tensor, with the nonzero component $\epsilon_T^{12}=1$. The symmetric traceless tensors $k_T^{i_1 ... i_n}$, constructed from the transverse momentum vector, are defined as:
\begin{align}
k_T^{ij}\equiv& k_T^i k_T^j +\frac{1}{2} \bm{k}_T^2 g_T^{ij}\,,\\
k_T^{ijk}\equiv& k_T^i k_T^j k_T^k +\frac{1}{4}\bm{k}_T^2 \left( g_T^{ij}k_T^k+g_T^{ik}k_T^j+g_T^{jk}k_T^i\right)\,,\\
k_T^{ijkl}\equiv& k_T^i k_T^j k_T^k k_T^l +\frac{1}{6} \bm{k}_T^2 \left(g_T^{ij}k_T^{kl}+g_T^{ik}k_T^{jl}+g_T^{il}k_T^{jk}\right.\notag\\
&\left.+g_T^{jk}k_T^{il}+g_T^{jl}k_T^{ik}+g_T^{kl}k_T^{ij}\right)\notag\\
&-\frac{1}{8} \bm{k}_T^4\left(g_T^{ij} g_T^{kl}+g_T^{ik} g_T^{jl}+g_T^{il} g_T^{jk}\right)\,,
\end{align}
which satisfy the tracelessness conditions
\begin{align}
    g_{Tij}k_T^{ij}=g_{Tij}k_T^{ijk}=g_{Tij}k_T^{ijkl}=0\,.
\end{align}

\begin{table}[H]
\centering
\caption{ Gluon TMDs classified by the polarization states
of the hadron (U: unpolarized; L or T: vector polarized; LL, LT, or TT:  tensor polarized) and the gluon (U: unpolarized; Circ: circularly polarized; Lin: linearly polarized), respectively.}\label{table:func}
    \setlength{\tabcolsep}{0.4cm}{
    \begin{tabular}{cccc}
    \hline
    $H \backslash g$ & U & Circ & Lin \\
    \hline
    U & $f_1$ &  & $h_1^\perp$ \\

    L &  & $g_1$ & $h_{1 L}^{\perp}$ \\

    T & $f_{1 T}^{\perp}$ & $g_{1 T}$ & $h_{1} \quad h_{1 T}^{\perp}$ \\

    LL&$f_{1LL}$&   &$h_{1LL}^\perp$\\

    LT&$f_{1LT}$&$g_{1LT}$&$h_{1LT} \quad h_{1LT}^\perp$\\

    TT&$f_{1TT}$&$g_{1TT}$&$h_{1TT} \quad h_{1TT}^{\perp} \quad h_{1TT}^{\perp \perp}$\\
    \hline
    \end{tabular}}
\end{table}

The gluon TMDs listed in Tab.~\ref{table:func}, as parametrized in Eqs.~\eqref{eq:phiU}-\eqref{eq:phiTT}, are classified by the polarization states of the hadron and the gluon. Among the full set of the nineteen functions, $f_{1T}^\perp$, $h_{1L}^\perp$, $h_1$, $h_{1T}^\perp$, $g_{1LT}$, and $g_{1TT}$ are T-odd, while the remaining distributions are T-even. In our previous work~\cite{Xie:2026uja}, we calculated the complete set of T-even gluon TMDs at tree-level within the spectator-model framework, with the spectator mass described by a flexible spectral function that samples a continuous range of physical values. However, tree-level amplitudes yield identically vanishing results for all T-odd gluon TMDs, as they lack the imaginary phases required to generate non-trivial T-odd observables. Nonvanishing T-odd TMDs only arise from the interference between two amplitudes with distinct imaginary phases, which necessitates the inclusion of loop-level effects, such as the exchange of an additional gluon, to generate the required complex phase.

\begin{figure}[htbp]
    \centering
    \includegraphics[width=0.8\columnwidth]{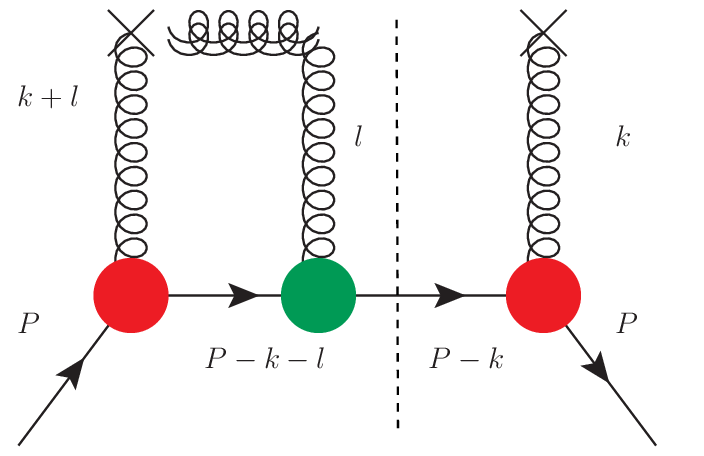}
    \caption{Diagram for one-loop order gluon-gluon correlator including the single-gluon exchange contribution. The double lines correspond to the eikonal propagator arising from the Wilson line in the operator definition of TMDs. The red blobs represent the deuteron-gluon-spectator vertex with color indices $ce$ and $ba$m respectively, while the green blob stands for the spectator-gluon-spectator vertex with color indices $edb$.}\label{fig:one-loop}     
\end{figure}

In this work, we generate the necessary phase interference by modeling the residual final-state interaction between the gluon and the spectator via the exchange of a soft gluon, as illustrated in Fig.~\ref{fig:one-loop}. The gluon-gluon correlator can be written as:
\begin{align}
&\Phi^{ij}\left(x,\bm{k}_T;S,T\right)\notag\\
=&\frac{1}{xP^+}\frac{1}{\left(2\pi\right)^3} \frac{1}{2\left(1-x\right)P^+}\epsilon^{\mu^\prime *}\left(P,\lambda\right)\epsilon^{\mu}\left(P,\lambda\right)\notag\\
&\times  G_{a a^\prime}^{i\beta*}\left(k,k\right) \mathcal{Y}^*_{\mu^\prime \nu^\prime \beta,a^\prime b^\prime}\left(P-k/2,k\right)\notag\\
&\times \epsilon_b^{*\nu}\left(P-k,\lambda_S\right) \epsilon_{b^\prime}^{\nu^\prime}\left(P-k,\lambda_S\right) (g_s n_-^\gamma f^{dac})\notag\\
&\times \int \frac{d^4 l}{(2 \pi)^4} \frac{-i \mathcal{X}_{\sigma_2 \nu \gamma}^{bde}(P-k-l/2,l)}{l^2-m_g^2}\frac{-i}{l^+ + i\varepsilon}\notag\\
&\times \frac{i \left(-g^{\sigma_1 \sigma_2}+(P-k-l)^{\sigma_1}(P-k-l)^{\sigma_2}/M_S^2\right)}{(P-k-l)^2-M_S^2+i \varepsilon}\notag\\
&\times G_{c c^\prime}^{j\alpha}\left(k,k+l\right) \mathcal{Y}_{\mu \sigma_1 \alpha,c^\prime e}(P-k/2-l/2,k+l)\,,\label{eq:phiWW}
\end{align}
where $\epsilon^{\mu}\left(P,\lambda\right)$ denotes the polarization vector of the deuteron, and $\epsilon_{b^\prime}^{\nu^\prime}\left(P-k,\lambda_S\right)$ represents the polarization vector of the spectator with color index $b^\prime$. Here, $g_s$ is the strong coupling constant. The Feynman rules for the eikonal gluon line and the eikonal vertex are given in detail in Ref.~\cite{Buffing:2017mqm}.

The structure of the gauge links in Eq.~\eqref{eq:phi0} is process dependent, with distinct Wilson line path topologies corresponding to different physical scattering processes. The gauge link structure encoded in the correlator of Eq.~\eqref{eq:phiWW} corresponds to final-state interactions between the spectator and the outgoing gluon, with a future-pointing closed path ([+,+]). The gluon TMDs extracted from this correlator are called Weizs$\ddot{a}$cker-Williams (WW) gluon TMDs, or $f$-type gluon TMDs, as the eikonal vertex Feynman rule involves the antisymmetric structure constants $f^{dac}$ of the color gauge group SU(3). 
In contrast, gauge links with a past-pointing closed path ([$-$,$-$]) correspond to initial-state interactions. Furthermore, gauge link paths, such as [+,$-$] and [$-$,+], incorporate both initial and final state interactions and give rise to dipole gluon TMDs (also called $d$-type gluon TMDs). For these distributions, the $f^{dac}$ structure constant in Eq.~\eqref{eq:phiWW} is replaced by $-i d^{dac}$, where $d^{dac}$ is the symmetric SU(3) structure constant.

The difference between correlators with future-pointing [+,+] and past-pointing [$-$,$-$] color paths (and between [+,$-$] and [$-$,+] paths) manifests as a sign flip of the $+i\varepsilon$ term in Eq.~\eqref{eq:phiWW}. Under this transformation, T-odd gluon TMDs change sign (i.e., [+,+]=$-$[$-$,$-$], [+,$-$]=$-$[$-$,+]). 
By contrast, T-even gluon TMDs are fully invariant under this transformation. Crucially, $f$-type and $d$-type gluon TMDs are physically distinct observables and cannot be directly related; they encode different aspects of gluonic information. Comprehensive discussions of gauge link topologies and their corresponding TMD observables can be found in Refs.~\cite{Boer:2016fqd,Bomhof:2006dp,Buffing:2013kca}.

The effective deuteron-gluon-spectator vertex $\mathcal{Y}^{\mu \nu \alpha}_{bc}$ is parametrized as
\begin{align}
&\mathcal{Y}_{bc}^{\mu \nu \alpha}\left(\overline{P},k\right)\notag\\
=&\Bigg\{2\left[-g^{\mu \nu}g_1\left(k^2\right)-\frac{2 \overline{P}^\mu \overline{P}^\nu}{M^2} g_3\left(k^2\right)\right]\overline{P}^\alpha\notag\\
&+2\left(\overline{P}^\nu g^{\alpha \mu}+\overline{P}^\mu g^{\alpha \nu}\right) g_2\left(k^2\right)\Bigg\}\delta_{bc}\,,\label{eq:vertex}
\end{align}
where $\overline{P}$ denotes the average momentum of the incoming and outgoing deuteron states, and $g_{1,2,3}\left(k^2\right)$ are the deuteron-gluon-spectator coupling form factors. This parametrization follows the conventional form factor decomposition for the vector current of a spin-1 particle~\cite{Arnold:1979cg,Brodsky:1992px,Zhang:2024nxl}. Consistent with our previous treatment of T-even TMDs in Ref.~\cite{Xie:2026uja}, we adopt an exponential form factor:
\begin{align}
g_{1,2,3}\left(k^2\right) =\kappa_{1,2,3} e^{k^2/\Lambda_S^2}\,,\label{eq:form}
\end{align}
where $\kappa_{1,2,3}$ are free coupling parameters and $\Lambda_S$ is a cut-off parameter introduced to regularize divergences arising from the point-like coupling.

The spectator-gluon-spectator vertex $\mathcal{X}_{\sigma_2 \nu \gamma}^{bde}$ is structurally analogous to the vertex in Eq.~\eqref{eq:vertex}, and contains both the antisymmetric $f^{bde}$ and symmetric $d^{bde}$ color structure constants:
\begin{align}
\mathcal{X}_{\sigma_2 \nu \gamma}^{bde}\left(\overline{P},k\right)=f^{bde}\mathcal{Y}_{\sigma_2 \nu \gamma}^f\left(\overline{P},k\right)-id^{bde} \mathcal{Y}_{\sigma_2 \nu \gamma}^d\left(\overline{P},k\right)\,,
\end{align}
In general, the vertices $\mathcal{Y}_{\sigma_2 \nu \gamma}^{f,d}$ are not required to be identical to the deuteron-gluon-spectator vertex $\mathcal{Y}_{\sigma_2 \nu \gamma}$ defined in Eq.~\eqref{eq:vertex}. For simplicity, however, we assume $\mathcal{Y}_{\sigma_2 \nu \gamma}^{f,d}$=$\mathcal{Y}_{\sigma_2 \nu \gamma}$. Under this assumption, the $f$-type and $d$-type T-odd gluon TMDs differ only by a constant color factor, given by:
\begin{align}
[+,+]/[+,-]=(f^{acd}f^{dca})/[(-id^{acd})(-id^{dca})]=9/5\,.
\end{align}
Accordingly, we present numerical results for the $f$-type gluon TMDs in the subsequent analysis.

The sum over all polarization states of the spectator is given by:
\begin{align}
&\sum_{\lambda_S}\epsilon_c^{*\nu}\left(P-k,\lambda_S\right) \epsilon_{c^\prime}^{\nu^\prime}\left(P-k,\lambda_S\right)\notag\\
=&-g^{\nu \nu^\prime}+\frac{\left(P-k\right)^\nu \left(P-k\right)^{\nu^\prime}}{M_S^2}\,.
\end{align}
And the term
\begin{align}
G_{ab}^{j\alpha}\left(p,k\right)=-\frac{ip^+}{k^2}\left(g^{j\alpha}-\frac{k^j n_-^\alpha}{p^+}\right)
\end{align}
corresponds to the Feynman rule for the field strength tensor $-i\left(p^\mu g^{\nu \rho}-p^\nu g^{\mu \rho}\right)\delta_{ab}$~\cite{Goeke:2006ef,Collins:2011zzd}. 

The six T-odd gluon TMDs listed in Tab.~\ref{table:func} are extracted via the appropriate projection of the gluon-gluon correlator $\Phi^{ij}\left(x,\bm{k}_T\right)$, with the explicit projection relations given by:
\begin{align}
f_{1T}^\perp \left(x,\bm{k}_T^2\right)=&\frac{M}{\epsilon_T^{k_T S_T}} g_T^{ij} \Phi_T^{ij}\left(x,\bm{k}_T\right)\,,\\
h_{1L}^\perp \left(x,\bm{k}_T^2\right)=&\frac{2M^2}{\bm{k}_T^4 S_L}\epsilon_{T \alpha}^{\left\{i\right.} k_T^{\left.j\right\}\alpha} \Phi_L^{ij}\left(x,\bm{k}_T\right)\,,\\
h_1 \left(x,\bm{k}_T^2\right)=&-\frac{M}{\epsilon_T^{k_T S_T}}\Phi_T^{ij}\left(x,\bm{k}_T\right)\notag\\
&\times\bigg(\frac{4}{\bm{k}_T^2}k_T^i k_T^j- \frac{k_T^{\left\{i\right.} S_T^{\left.j\right\}}}{\bm{k}_T \cdot \bm{S}_T}-3g_T^{ij}\bigg)\,,\\
h_{1T}^\perp \left(x,\bm{k}_T^2\right)=&-\frac{2M^3}{\bm{k}_T^2 \epsilon_T^{k_T S_T}}\Phi_T^{ij}\left(x,\bm{k}_T\right)\notag\\
&\times\left(- \frac{k_T^{\left\{i\right.} S_T^{\left.j\right\}}}{\bm{k}_T \cdot \bm{S}_T}-g_T^{ij}\right)\,,\\
g_{1LT} \left(x,\bm{k}_T^2\right)=&\frac{M}{\epsilon_T^{k_T S_{LT}}} i\epsilon_T^{ij} \Phi_{LT}^{ij}\left(x,\bm{k}_T\right)\,,\\
g_{1TT} \left(x,\bm{k}_T^2\right)=&\frac{-M^2}{\epsilon_{T\gamma}^{\beta} k_T^{\gamma \alpha} S_{TT\alpha \beta}} i\epsilon_T^{ij} \Phi_{TT}^{ij}\left(x,\bm{k}_T\right)\,.
\end{align}

To generate the nonvanishing imaginary phase required for non-trivial T-odd gluon TMDs, we apply the Cutkosky rules to the one-loop diagram, accounting for cuts through both the eikonal propagator and the spectator propagator inside the loop. This procedure is implemented via the following replacements for the propagators:
\begin{align}
\frac{1}{l^+ + i\epsilon} &\to -2\pi i \delta(l^+)\,,\\
\frac{1}{(P-k-l)^2-M_S^2+i\epsilon} &\to -2\pi i \delta((P-k-l)^2-M_S^2) \,.
\end{align}
With the spectator required to be on-shell, satisfying $\left(P-k\right)^2=M_S^2$, the virtuality of the gluon takes the form:
\begin{align}
k^2=-\frac{\bm{k}_T^2+L^2_S}{1-x}\,,
\end{align}
where $L^2_S=xM_S^2-x(1-x)M^2$. This on-shell relation holds analogously for other loop momenta, including $l$ and $k+l$, and is applied consistently to evaluate all the propagators and form factors in the remainder of the calculation. 

All six T-odd gluon TMDs admit a universal integral structure, denoted generically as $F\left(x,\bm{k}_T^2\right)$, given by:
\begin{align}
F\left(x,\bm{k}_T^2\right)=&\int \frac{d^2 \bm{l}_T}{(2\pi)^2} \frac{e^{-\frac{\bm{k}_T^2+L_S^2}{(1-x)\Lambda_S^2}}}{\bm{k}_T^2+L_S^2}\frac{e^{-\frac{\bm{l}_T^2+L_S^2}{(1-x)\Lambda_S^2}}}{\bm{l}_T^2+L_S^2}\notag\\
&\times\frac{e^{-\frac{(\bm{k}_T+\bm{l}_T)^2+L_S^2}{(1-x)\Lambda_S^2}}}{(\bm{k}_T+\bm{l}_T)^2+L_S^2}\notag\\
&\times \sum_{i,j,k}^{1,2,3} \mathcal{C}_{ijk}^{[F]}\left(x,\bm{k}_T^2\right) g_s \kappa_i \kappa_j \kappa_k\,,\label{eq:F}
\end{align}
where $\kappa_{i,j,k}$ are the coupling parameters associated with the exponential form factors defined in Eq.~\eqref{eq:form}. The full analytic expressions for the coefficients $\mathcal{C}_{ijk}^{[F]}$ are compiled in Appendix~\ref{appendix1}.

In the limiting case where the coupling parameter $\kappa_2$ is set to zero (referred to as the $g_2$-vanishing approximation), the T-odd TMDs $g_{1LT}$ and $g_{1TT}$ vanish identically. The remaining nonvanishing T-odd distributions $f_{1T}^\perp$, $h_1$ and $h_{1T}^\perp$ satisfy the following model-dependent relations:
\begin{align}
f_{1T}^{\perp(g_{1,3})}=\frac{1}{5} h_1^{(g_{1,3})}=-\frac{\bm{k}_T^2}{2M^2}h_{1T}^{\perp(g_{1,3})}\,.\label{eq:g2}
\end{align}

To incorporate the effects of $q\bar{q}$ contributions and model the continuous kinematic range of the spectator mass $M_S$, we adopt a spectral function $\rho\left(M_S\right)$ with the following parametrization~\cite{Bacchetta:2020vty}:
\begin{align}
\rho\left(M_S\right)=\mu^{2a}\left[\frac{A}{B+\mu^{2b}}+\frac{C}{\pi \sigma} e^{-\frac{(M_S-D)^2}{\sigma^2}}\right]\,,
\end{align}
where $\mu^2=M_S^2-M^2$, and the set of free parameters is given by $\{X\}\equiv\{A,B,a,b,C,D,\sigma\}$. This parametrization combines a smooth non-resonant background with a Gaussian resonant peak, capturing the full continuum of spectator hadronic states. The gluon TMDs are then obtained by integrating over the spectator mass weighted by the spectral function $\rho\left(M_S\right)$:
\begin{align}
F\left(x,\bm{k}_T^2\right)=\int_{M}^{\infty}dM_S\ \rho\left(M_S\right) F\left(x,\bm{k}_T^2;M_S\right)\,.\label{eq:weighted}
\end{align}
This spectral weighting procedure effectively accounts for the full continuum of spectator hadronic configurations, significantly extending the physical applicability of the model framework.

\section{Numerical results}\label{section3}

In this section, we present the numerical results for all six leading-twist, $f$-type T-odd gluon TMDs listed in Tab.~\ref{table:func}. Under the assumption that the form factors $g_{1,2,3}\left(k^2\right)$ for both $f$-type and $d$-type distributions are identical to those used in our T-even TMD analysis, the T-odd $d$-type gluon TMDs in our model are related to their $f$-type counterparts solely by the constant color factor derived earlier. The model parameters, previously determined from a fit to the nNNPDF1.0 parametrization for the integrated T-even gluon unpolarized TMD $f_1 (x)$ at the low scale $Q_0=2~\mathrm{GeV}$ in Ref.~\cite{Xie:2026uja}, are listed in Tab.~\ref{table:parm}. 

To enable a direct quantitative comparison and consistency check against theoretical constraints, we analyze the $n$-th transverse moment $F^{(n)}\left(x\right)$ of the TMDs, defined as:
\begin{align}
F^{(n)}\left(x\right)=\int d^2 \bm{k}_T \left(\frac{\bm{k}_T^2}{2M^2}\right)^{(n)} F\left(x,\bm{k}_T^2\right)\,.
\end{align} 
For the T-odd distributions, the moment order $n$ is chosen to match that of the T-even functions appearing in the corresponding model-independent positivity bounds listed in Appendix~\ref{appendix2}. For instance, we display the first transverse moment ($n=1$) of T-odd TMDs $f_{1T}^\perp$ and $g_{1LT}$, as the first transverse moments of the T-even TMDs $f_{1LT}^{(1)}$, $g_{1T}^{(1)}$, and $h_{1LT}^{(1)}$ were shown in our prior T-even analysis~\cite{Xie:2026uja}, with all of these distributions constrained by the positivity bound in Eq.~\eqref{eq:bound1}. Conversely, for the bound in Eq.~\eqref{eq:bound2}, we present the T-odd TMD $h_{1T}^\perp$ and the T-even TMD $h_{1LT}^\perp$ via their second moment ($n=2$) $F^{(2)}\left(x\right)$. This choice also allows us to explicitly verify the specific relation derived in Eq.~\eqref{eq:g2}.

\begin{table}[H]
\centering
\caption{Central column:mean values and uncertainties of the fitted model parameters taking from Ref.~\cite{Xie:2026uja}. Rightmost column: corresponding values for replica 60}\label{table:parm}
    \setlength{\tabcolsep}{0.4cm}{
    \begin{tabular}{ccc}
    \hline
    Parameter & Mean & Replica 60 \\
    \hline
    $\kappa_1$ &0.713 $\pm$ 0.604 &0.350 \\
    $\kappa_2$ &0.334 $\pm$ 0.303 &0.149 \\
    $\kappa_3$ &15.56 $\pm$ 6.06 &12.88 \\
    $\Lambda_S$&1.34 $\pm$ 0.18 &1.36 \\
    $a$&1.564 $\pm$ 1.442 &1.529 \\
    $b$&10.14 $\pm$ 5.66 &5.93 \\
    $A$&138 $\pm$ 141 &221 \\
    $B$&5.86 $\pm$ 6.65 &6.42 \\
    $C$&305 $\pm$ 164 &359 \\
    $D$&1.19 $\pm$ 0.55 &1.31 \\
    $\sigma$&0.683 $\pm$ 0.226 &0.674 \\
    \hline
    \end{tabular}}
\end{table}

\begin{figure*}[htbp]
    \centering
    \includegraphics[width=1.8\columnwidth]{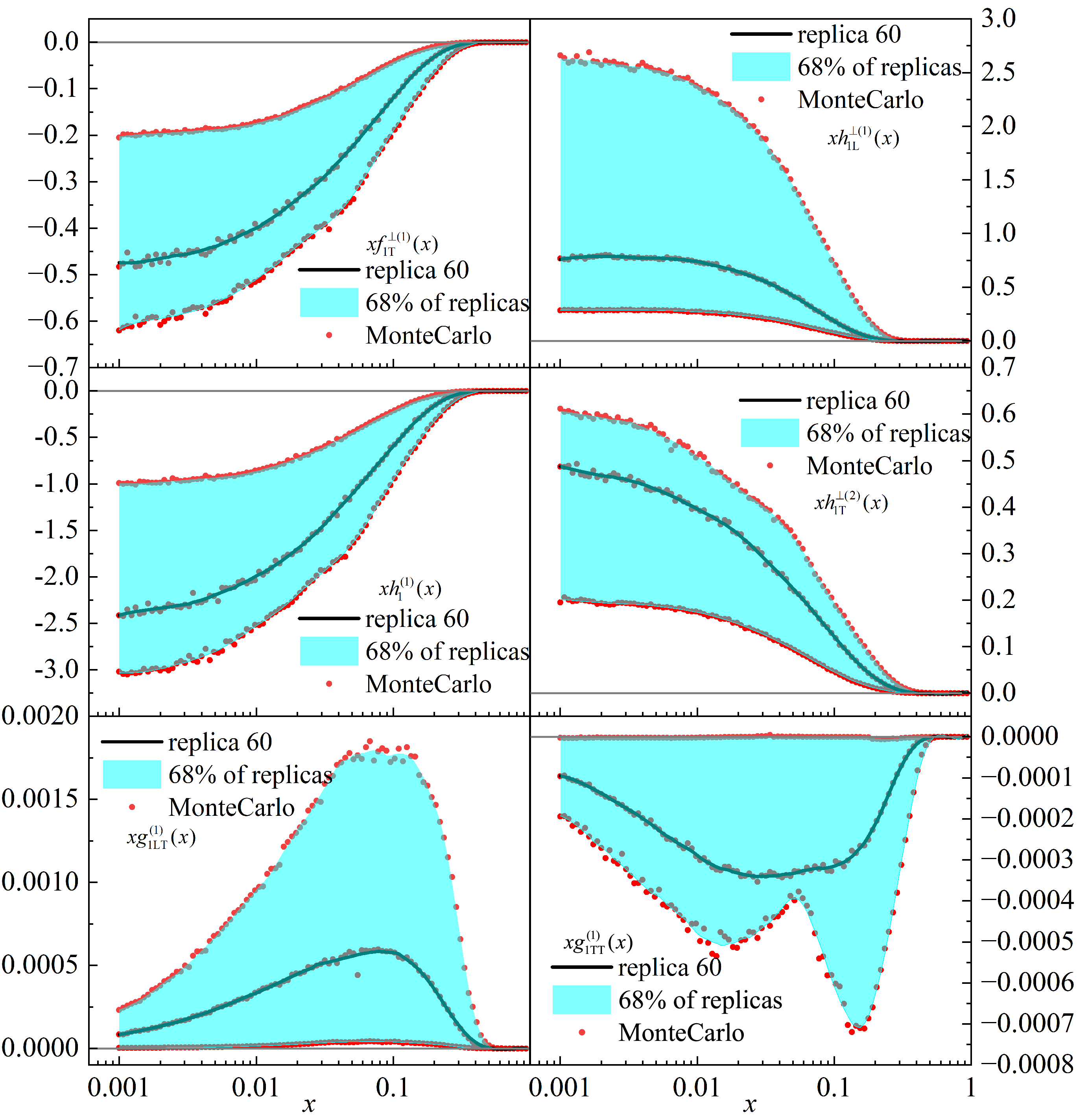}
    \caption{T-odd gluon TMDs $xf_{1T}^{\perp (1)}$(upper-left), $xh_{1L}^{\perp (1)}$(upper-right), $xh_{1}^{(1)}$(mid-left), $xh_{1T}^{\perp (2)}$(mide-right), $xg_{1LT}^{(1)}$(lower-left), and $xg_{1TT}^{(1)}$(lower-right) as functions of $x$ at $Q_0=2~\mathrm{GeV}$. 
    The cyan band shows the 68\% uncertainty band, the solid line corresponds to the result from replica 60, and red points mark the Monte Carlo integration outputs. 
    \label{fig:x} }   
\end{figure*}
\begin{figure*}[htbp]
    \centering
    \includegraphics[width=1.8\columnwidth]{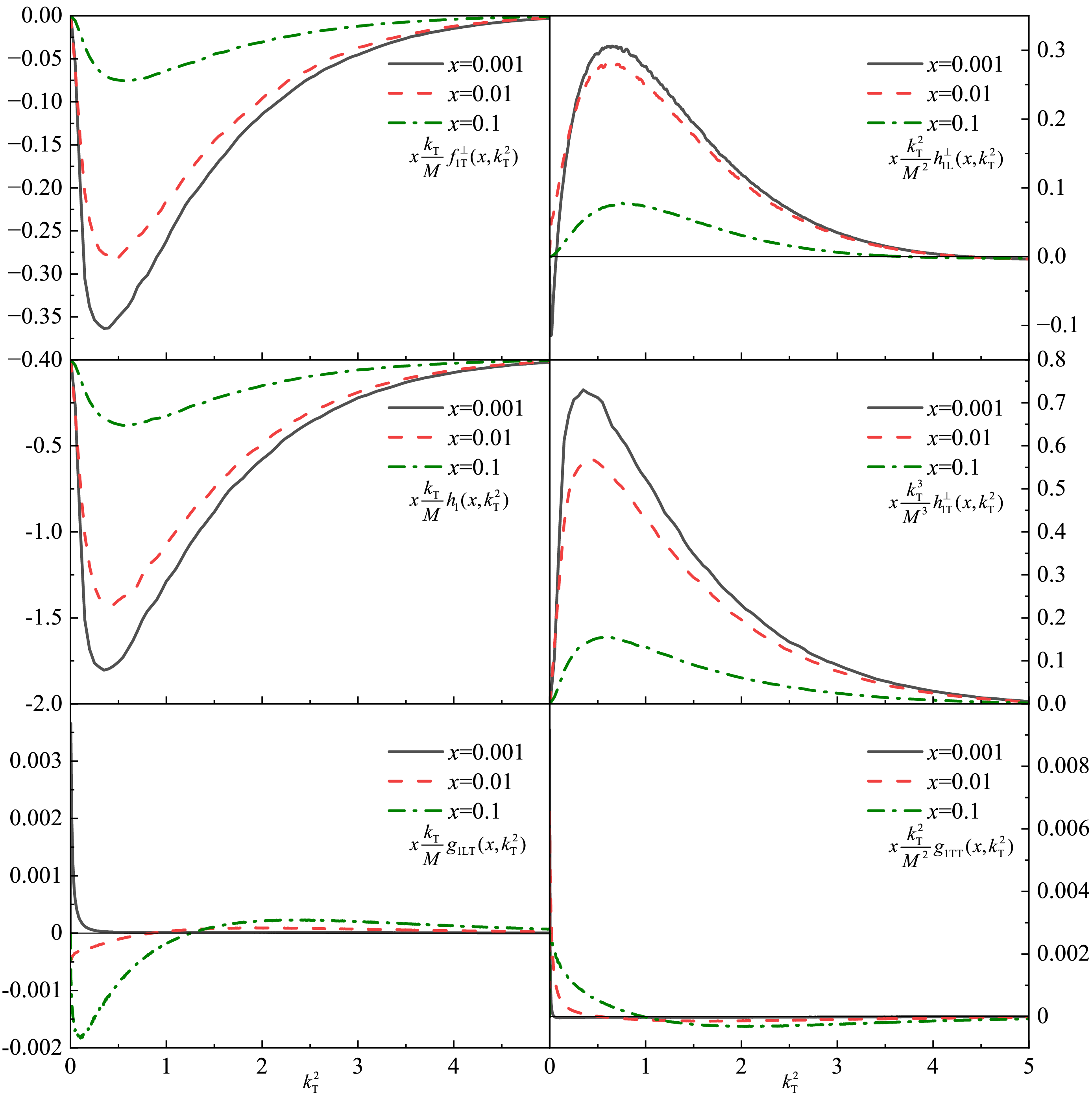}
    \caption{T-odd gluon TMDs $x\bm{k}_T /M~f_{1T}^{\perp}$(upper-left), $x\bm{k}_T^2 /M^2~h_{1L}^{\perp}$(upper-right), $x\bm{k}_T /M~h_{1}$(mid-left), $x\bm{k}_T^3 /M^3~h_{1T}^{\perp}$(mide-right), $x\bm{k}_T /M~g_{1LT}$(lower-left), and $x\bm{k}_T^2 /M^2~g_{1TT}$(lower-right) as functions (replica 60) of $\bm{k}_T^2$ at $Q_0=2~\mathrm{GeV}$. Solid, dashed, and dash-dotted lines represent the cases $x=0.001$, 0.02 and 0.1,  respectively. \label{fig:kt} }   
\end{figure*}

Fig.~\ref{fig:x} shows the numerical results for the first transverse moments of the T-odd gluon TMDs $xf_{1T}^{\perp (1)}$, $xh_{1L}^{\perp (1)}$, $xh_{1}^{(1)}$, $xh_{1T}^{\perp (2)}$, $xg_{1LT}^{(1)}$, and $xg_{1TT}^{(1)}$ as functions of $x$ at the scale $Q_0=2~\mathrm{GeV}$. 
The solid black line corresponds to the result of the most representative replica 60. The uncertainty band is constructed by excluding the largest and smallest 16\% of all 100 replicas, roughly corresponding to $1\sigma$ standard deviation. Given the computational complexity of the four-dimensional integrations over the transverse momenta $\bm{l}_T$ and $\bm{k}_T$, we perform all numerical integrations using a Monte Carlo algorithm, with the resulting discrete points shown as red markers.  The strong coupling constant is fixed to $g_s=\sqrt{4\pi \alpha_s(Q_0)}=\sqrt{1.2\pi}$.

Our numerical results show that $xf_{1T}^{\perp (1)}$, $xh_{1}^{(1)}$, and $xg_{1TT}^{(1)}$ are negative in sign over the full range $0.001<x<1$, while $xh_{1L}^{\perp (1)}$, $xh_{1T}^{\perp (2)}$, and $xg_{1LT}^{(1)}$ are positive in sign. The magnitudes of $xf_{1T}^{\perp (1)}$ and $xh_{1}^{(1)}$ rise smoothly with increasing $x$, whereas $xh_{1L}^{\perp (1)}$ and $xh_{1T}^{\perp (2)}$ decrease as $x$ increases. We find that $xf_{1T}^{\perp (1)}$, $xh_{1}^{(1)}$ and $xh_{1T}^{\perp (2)}$ approximately satisfy the relation in Eq.~\eqref{eq:g2}, consistent with the $g_2$-vanishing approximation defined earlier. Notably, the magnitudes of $xg_{1LT}^{(1)}$ and $xg_{1TT}^{(1)}$ are substantially smaller than those of the other T-odd TMDs; this feature is highly sensitive to the $g_2\left(k^2\right)$ form factor, as these distributions vanish identically in the $g_2$-vanishing limit. 

In Fig.~\ref{fig:kt}, we show the $\bm{k}_T^2$-dependence of $x\bm{k}_T^n/M^n\,F\left(x,\bm{k}_T^2\right)$ for all six T-odd gluon TMDs at $x=10^{-3}$ (solid), $x=10^{-2}$ (dashed), and $x=10^{-1}$ (dash-dotted). All curves are obtained using parameters from replica 60 at the scale $Q_0=2~\mathrm{GeV}$. The $\bm{k}_T^2$-dependence of these distributions deviates significantly from a simple Gaussian form, instead exhibiting a prominent flattened tail at large $\bm{k}_T^2$.

For the T-odd gluon TMDs associated with a vector polarized deuteron, all distributions exhibit a peak in the region $\bm{k}_T^2<1~\mathrm{GeV}^2$, with the peak position shifting to larger $\bm{k}_T^2$ values as $x$ increases.  Strikingly, $x\bm{k}_T^2/M^2~h_{1L}^\perp$ at $x=10^{-3}$ displays a long tail and changes sign, with a node at $\bm{k}_T^2 \approx 0.07~\mathrm{GeV}^2$.

For the tensor polarized case, the magnitudes of $x\bm{k}_T/M~g_{1LT}$ and $x\bm{k}_T^2/M^2~g_{1TT}$ are several times smaller than those of the vector-polarized T-odd TMDs. The $\bm{k}_T^2$-profile of $x\bm{k}_T/M~g_{1LT}$ also shows strong $x$-dependence, with substantial shape variations across the three $x$ values considered. Intriguingly, both $x\bm{k}_T/M~g_{1LT}$ and $x\bm{k}_T^2/M^2~g_{1TT}$ develop a node in the region $\bm{k}_T^2 >1~\mathrm{GeV}^2$ for all values of $x$ values considered.

Finally, we have verified that the numerical results for all T-odd gluon TMDs in our model satisfy the model-independent positivity bounds listed in Appendix~\ref{appendix2}. These bounds provide crucial constraints for both TMD and collinear distributions, as discussed in detail in Ref.~\cite{Cotogno:2017puy}.

\section{Conclusion}\label{section4}

In this paper, we perform a comprehensive spectator-model calculation of all leading-twist T-odd gluon TMDs for the tensor-polarized deuteron. The model assumes that an on-shell deuteron can emit a timelike off-shell gluon, with the residual system approximated as a single on-shell spectator whose mass is described by a continuous spectral function. Describing spin-1 deuteron polarization demands both a spin vector $S^\mu$ and symmetric traceless spin tensor $T^{\mu\nu}$. We construct the deuteron–gluon–spectator vertex via Lorentz-invariant vector-current form factors and adopt the exponential coupling form factor.

To generate  nonvanishing T-odd distributions, we incorporated final-state interactions via single-gluon exchange between the struck outgoing gluon and the spectator system.   
Due to the presence of the Wilson line in the operator definition of the gluon TMDs, two primary classes of gluon TMDs: the Weizs$\ddot{a}$cker-Williams ($f$-type) TMDs and the dipole ($d$-type) TMDs emerge. In principle, the $f$-type and $d$-type gluon TMDs are physically distinct observables and cannot be directly related. For the sake of simplicity, however, our model assumes that their differences are encapsulated solely in the color structure of the spectator-gluon-spectator vertex. Under this assumption, the magnitude of the $d$-type gluon TMDs is related to their $f$-type counterparts by a constant factor of $5/9$.

We provide full analytical expressions for all six T-odd gluon TMDs, along with numerical results for the $f$-type distributions. All model parameters are kept fully consistent with those in our earlier T-even gluon TMDs analysis, which were fixed by fitting the nNNPDF1.0 parametrization for the integrated unpolarized gluon TMD $f_1\left(x\right)$ at the scale $Q_0=2~\mathrm{GeV}$. Our numerical results show that the magnitudes of $xg_{1LT}^{(1)}$ and $xg_{1TT}^{(1)}$ are substantially smaller than those of the other T-odd TMDs; this feature is highly sensitive to the $g_2\left(k^2\right)$ form factor, and warrants dedicated further investigation. Furthermore, we have verified that our model results satisfy the relevant positivity bounds for both T-odd gluon TMDs and their collinear counterparts. Our analysis indicates that the magnitudes of T-odd gluon TMDs could be significant, providing valuable input for future experimental measurements and phenomenological model developments. 

The flexibility of our spectator-model framework enables a range of future extensions to incorporate more complex coupling structure and color interactions in the vertices. Such improvements will allow for a more rigorous treatment of the differences between $f$-type and $d$-type gluon TMDs, beyond the simple constant color factor approximation adopted in this work. We anticipate that our results will provide a critical baseline for interpreting data from ongoing and future experiments probing spin-1 hadron structure, such as those at the Thomas Jefferson National Accelerator Facility (JLab)~\cite{Maxwell:2018gci}, the Fermi National Accelerator Laboratory (Fermilab)~\cite{Keller:2022abm}, the nuclotron-based ion collider facility (NICA)~\cite{Arbuzov:2020cqg}, LHC(Large Hadron Collider)-spin~\cite{Aidala:2019pit}, and electron-ion colliders (EIC, EicC)~\cite{AbdulKhalek:2021gbh,Anderle:2021wcy}.

\section*{Acknowledgements}
This work is partially supported by the National Natural Science Foundation of China under grant number 12150013. Xiupeng Xie is also supported by the SEU Innovation Capability Enhancement Plan for Doctoral Students under grant number CXJH$\_$SEU 25138.

\begin{widetext}
\appendix
\section{Full expressions of T-odd gluon TMDs}\label{appendix1}
In the following, we list the final expressions of the $\mathcal{C}_{ijk}^{[F]}$ coefficients in Eq.~\eqref{eq:F} for each T-odd gluon TMDs and for $i,j,k=1,2,3$.

\subsection{Sivers function \texorpdfstring{$f_{1T}^\perp$}{f1T}}
\begin{align}
&\mathcal{C}_{111}^{[f_{1T}^\perp]}\notag\\
=&3 (1-x) \left(l_T\cdot k_T+k_T^2\right) \left(l_T^2+2M_S^2\right) \left[l_T\cdot k_T \left(k_T^2+M^2(x-1)^2-M_S^2\right)+k_T^2 l_T^2\right]\left(4 \pi ^3 x k_T^2 M_S^4\right)^{-1}\,,\label{f1T:111} \\
&\mathcal{C}_{222}^{[f_{1T}^\perp]}\notag\\
=&-3 \Bigg[k_T^2 l_T^2 \left(l_T^2 \left(-2 (x-1) k_T^2-M^2 x(x-1)^2+x (2 x-1) M_S^2\right)+M_S^2 \left(((x-2) x+2) k_T^2+(x-2)x \left(M^2 (x-1)^2-M_S^2\right)\right)\right)\notag\\
&+\left(l_T\cdot k_T\right){}^2 \left(2 l_T^2 \left((1-2 x) k_T^2-M^2 x (x-1)^2+2 xM_S^2\right)+M_S^2 \left(((x-6) x+6) k_T^2+(x-2) x \left(M^2(x-1)^2-M_S^2\right)\right)\right)\notag\\
&+l_T\cdot k_T \Big(k_T^2\left(l_T^2 \left((3 (x-1) x+2) M_S^2-3 M^2 (x-1)^2 x\right)+(2-3x) l_T^4+2 (x-2) x M_S^2 \left(M^2(x-1)^2-M_S^2\right)\right)\notag\\
&+k_T^4 \left((2-3 x) l_T^2+(x-2)^2M_S^2\right)+x \left(M^2 (x-1)^2-M_S^2\right) \left(M_S^2\left((x-2) l_T^2+M^2 x (x-1)^2\right)-l_T^4-xM_S^4\right)\Big)\notag\\
&+2 (x+1) M_S^2 \left(l_T\cdot k_T\right){}^3\Bigg]\left(16 \pi ^3 x k_T^2 M_S^4\right)^{-1}\,,\\
&\mathcal{C}_{333}^{[f_{1T}^\perp]}\notag\\
=&3 \left(l_T\cdot k_T+k_T^2\right) \left(l_T\cdot k_T\left(k_T^2+M^2 (x-1)^2-M_S^2\right)+k_T^2 l_T^2\right)\left(l_T^2 \left(k_T^2+M^2 (x-1)^2-M_S^2\right)-4 M_S^2 l_T\cdot k_T\right)\notag\\
&\times \left(l_T^2 \left(2 l_T\cdot k_T+k_T^2+M^2 (x-1)^2+3M_S^2\right)+4 M_S^2 l_T\cdot k_T+l_T^4\right)\left(256 \pi ^3 M^6(1-x) x k_T^2 M_S^4\right)^{-1}\,,\label{f1T:333}
\end{align}
\begin{align}
&\mathcal{C}_{112}^{[f_{1T}^\perp]}+\mathcal{C}_{121}^{[f_{1T}^\perp]}+\mathcal{C}_{211}^{[f_{1T}^\perp]}\notag\\
=&3 \Bigg[2 (x-1) k_T^2 l_T^2 \left(2 M_S^2 \left(3 k_T^2-xl_T^2+M^2 (x-1) x\right)+l_T^2 \left(5 k_T^2+M^2 (x-1) x\right)-2x M_S^4\right)+2 \left(l_T\cdot k_T\right){}^2 \notag\\
&\times\left(M_S^2\left((4 x-6) k_T^2+(2 x-1) \left(2 M^2 (x-1)^2-3l_T^2\right)\right)+l_T^2 \left((6 x-5) k_T^2+M^2 (4 x-3)(x-1)^2\right)+2 (2 (x-2) x+1) M_S^4\right)\notag\\
&+l_T\cdot k_T\Big(l_T^4 \left((11 x-10) k_T^2+M^2 x (x-1)^2-xM_S^2\right)+l_T^2 (-2 (2 x-3) (x-1) M_S^2 \left(k_T^2+M^2(x-1) x\right)\notag\\
&+2 M^2 (5 x-3) (x-1)^2 k_T^2+(11 x-10) k_T^4-M^4 x(x-1)^4+x (4 x-5) M_S^4)+2 M_S^2 \big(2 k_T^2 (M^2 (3x-1) (x-1)^2\notag\\
&+((x-3) x+1) M_S^2)+(5 x-6) k_T^4+x \left(M^4(x-1)^4-2 M^2 (x-1)^3 M_S^2+(2 x-3) M_S^4\right)\big)\Big)\notag\\
&-8 xM_S^2 \left(l_T\cdot k_T\right){}^3\Bigg]\left(16 \pi ^3 x k_T^2M_S^4\right)^{-1}\,,
\end{align}
\begin{align}
&\mathcal{C}_{113}^{[f_{1T}^\perp]}+\mathcal{C}_{131}^{[f_{1T}^\perp]}+\mathcal{C}_{311}^{[f_{1T}^\perp]}\notag\\
=&3 l_T^2 \left(l_T\cdot k_T+k_T^2\right) \left(2 l_T\cdot k_T+2k_T^2-(x-2) l_T^2+2 M^2 (x-1)^2+(6-4 x) M_S^2\right)\notag\\
&\times \left(l_T\cdot k_T \left(k_T^2+M^2 (x-1)^2-M_S^2\right)+k_T^2l_T^2\right)\left(16 \pi ^3 M^2 x k_T^2 M_S^4\right)^{-1}\,,\label{f1T:113}
\end{align}
\begin{align}
&\mathcal{C}_{223}^{[f_{1T}^\perp]}+\mathcal{C}_{232}^{[f_{1T}^\perp]}+\mathcal{C}_{322}^{[f_{1T}^\perp]}\notag\\
=&3 \Bigg[4 \left(l_T\cdot k_T\right){}^3 \bigg(M_S^2\left(\left(x^2+2\right) k_T^2+(2-(x-2) x) l_T^2+M^2 (x-2) x(x-1)^2\right)-x l_T^2 \left(k_T^2+M^2 (x-1)^2\right)\notag\\
&+(6-5 x) xM_S^4\bigg)+l_T^2 l_T\cdot k_T \Big(k_T^4 \left((x (3 x-19)+10)l_T^2-M^2 (9 x-4) (x-1)^2+(x (-2 (x-6) x-15)+4) M_S^2\right)\notag\\
&+k_T^2(l_T^2 \left(M^2 (x-4) (3 x-1) (x-1)^2+(4-(x-3) x (6 x-5))M_S^2\right)+(x-2) (3 x-2) l_T^4-M^4 (3 x-2) (x-1)^4\notag\\
&-2 M^2 (x (2(x-4) x+5)+2) (x-1)^2 M_S^2+\left(x \left(-4 x^3+8x-3\right)+2\right) M_S^4)+(2-5 x) k_T^6+x \left(M^2(x-1)^2-M_S^2\right)\notag\\
&\times \Big(-2 (x-2)^2 l_T^2 M_S^2+(x-2) l_T^4+M^4(x-1)^4-2 M^2 (x-1)^4 M_S^2+(1-2 x (x (2 x-3)+2))M_S^4\Big)\Big)\notag\\
&-k_T^2 l_T^4 \bigg(k_T^2 \left(((7-2 x) x-4)l_T^2+2 (x-1)^2 \left(M^2 (3 x-2)+(x-2) M_S^2\right)\right)+(5x-4) k_T^4+x \Big(l_T^2 ((x (4 x-9)+4) M_S^2\notag\\
&-M^2 (x-2)(x-1)^2)+M^4 (x-1)^4+2 M^2 (x-3) (x-1)^3 M_S^2+(2 x (x (2x-1)-2)+1) M_S^4\Big)\bigg)+2 l_T^2 \left(l_T\cdot k_T\right){}^2\notag\\
&\times \bigg(k_T^2 \left((x-3) (2 x-1) l_T^2-2 M^2 (3 x-1)(x-1)^2-(x ((x-12) x+10)-6) M_S^2\right)+(1-5 x) k_T^4\notag\\
&-M_S^2\left(3 (x-2) x l_T^2+M^2 (x ((x-6) x+6)+2) (x-1)^2\right)+M^2(x-3) x (x-1)^2 l_T^2-M^4 (x-1)^5\notag\\
&-(x (x (x+8)-11)-1)M_S^4\bigg)+8 M_S^2 \left(l_T\cdot k_T\right){}^4\Bigg]\left(64 \pi^3 M^2 (1-x) x k_T^2 M_S^4\right)^{-1}\,,
\end{align}
\begin{align}
&\mathcal{C}_{221}^{[f_{1T}^\perp]}+\mathcal{C}_{212}^{[f_{1T}^\perp]}+\mathcal{C}_{122}^{[f_{1T}^\perp]}\notag\\
=&3 \Bigg[k_T^2 l_T^2 \bigg(l_T^2 \left(-8 (x-1) k_T^2-3 M^2 x(x-1)^2+x (6 x-5) M_S^2\right)+2 M_S^2 \Big((x-2)^2 k_T^2+M^2(x-3) x (x-1)^2\notag\\
&+x (2 (x-1) x+1) M_S^2\Big)\bigg)-2\left(l_T\cdot k_T\right){}^2 \bigg(l_T^2 \left((6 x-4) k_T^2+M^2(3 x-1) (x-1)^2+(1-6 x) M_S^2\right)\notag\\
&-(x-2) M_S^2 \left((x-4)k_T^2+M^2 x (x-1)^2\right)+x (3 x-4) M_S^4\bigg)+l_T\cdot k_T\bigg(-2 l_T^4 \left((5 x-4) k_T^2+M^2 x (x-1)^2-xM_S^2\right)\notag\\
&+l_T^2 \bigg(k_T^2 \left((x (8 x-11)+6) M_S^2-M^2(x-1)^2 (9 x-2)\right)+(8-10 x) k_T^4+x \Big(M^4 (x-1)^4+2 M^2 (2x-3) (x-1)^2 M_S^2\notag\\
&+(5-4 x) M_S^4\Big)\bigg)+2 M_S^2 \bigg(xk_T^2 \left(M^2 (2 x-5) (x-1)^2+(x (2 x-5)+5) M_S^2\right)+(x-3)(x-2) k_T^4\notag\\
&+(x-1) x^2 \left(M^4 (x-1)^3+M^2 (x-1) (2 x-3) M_S^2-2M_S^4\right)\bigg)\bigg)+4 (2 x+1) M_S^2 \left(l_T\cdot k_T\right){}^3\Bigg]\left(16 \pi ^3 x k_T^2 M_S^4\right)^{-1}\,,
\end{align}
\begin{align}
&\mathcal{C}_{331}^{[f_{1T}^\perp]}+\mathcal{C}_{313}^{[f_{1T}^\perp]}+\mathcal{C}_{133}^{[f_{1T}^\perp]}\notag\\
=&3 \left(l_T\cdot k_T+k_T^2\right) \left(l_T\cdot k_T\left(k_T^2+M^2 (x-1)^2-M_S^2\right)+k_T^2 l_T^2\right)\Bigg[l_T^4 \Big(-2 (x-1) l_T\cdot k_T-(2 x-3) \left(k_T^2+M^2(x-1)^2\right)\notag\\
&+(3-4 x) M_S^2\Big)+l_T^2 \bigg(2 l_T\cdot k_T\left(k_T^2+M^2 (x-1)^2-(2 x+1) M_S^2\right)+\left(k_T^2+M^2(x-1)^2+M_S^2\right)\notag\\
&\times \left(k_T^2+M^2 (x-1)^2+(3-4 x)M_S^2\right)\bigg)+4 M_S^4 \left(-l_T\cdot k_T-k_T^2+M^2(x-1)^2\right)-2 M_S^2 \Big(2 l_T\cdot k_T \left(k_T^2+M^2(x-1)^2\right)\notag\\
&+4 \left(l_T\cdot k_T\right){}^2+\left(k_T^2+M^2(x-1)^2\right){}^2\Big)-\left((x-1) l_T^6\right)-2M_S^6\Bigg]\left(64 \pi ^3 M^4 (1-x) x k_T^2 M_S^4\right)^{-1}\,,\label{f1T:133}
\end{align}

\begin{align}
&\mathcal{C}_{332}^{[f_{1T}^\perp]}+\mathcal{C}_{323}^{[f_{1T}^\perp]}+\mathcal{C}_{233}^{[f_{1T}^\perp]}\notag\\
=& 3 \bigg\{ -16 M_S^2 \big( M^2 (x-1)^2 + k_T^2 - l_T^2 - 3 M_S^2 \big) (l_T \cdot k_T)^4 \notag \\
& + 4 \big[ 4 (3-2x) x M_S^6 + 4 \big( M^2 x (2x-3) (x-1)^2 + (x(2x-1)+3) k_T^2 \big) M_S^4 - 4 k_T^2 \big( M^2 (x-1)^2 + k_T^2 \big) M_S^2 \notag \\
& + l_T^4 \big( -M^2 x (x-1)^2 - x k_T^2 + (3x+2) M_S^2 \big) + 2 l_T^2 \big( (6-(x-4)x) M_S^4 + \big( M^2 (x-3) (x+1) (x-1)^2 \notag \\
& + (x^2-3) k_T^2 \big) M_S^2 + \big( M^2 (x-1)^2 + k_T^2 \big)^2 \big) \big] (l_T \cdot k_T)^3 + 2 l_T^2 \big[ 6 k_T^6 + \big( 14 M^2 (x-1)^2 \notag \\
& + (7-5x) l_T^2 - 22 M_S^2 \big) k_T^4 + 2 \big( 5 M^4 (x-1)^4 - M^2 (3x-5) l_T^2 (x-1)^2 + (1-2x) l_T^4 \notag \\
& + (8x(2x-1)+11) M_S^4 - \big( 8 M^2 (x-1)^2 + (-5x^2+x+5) l_T^2 \big) M_S^2 \big) k_T^2 + 2 x l_T^4 \big( 2 M_S^2 - M^2 (x-1)^2 \big) \notag \\
& + l_T^2 \big( -M^4 (x-3) (x-1)^4 + 2 M^2 (x(3x-5)-3) M_S^2 (x-1)^2 + ((19-6x)x+3) M_S^4 \big) \notag \\
& + 2 \big( M^6 (x-1)^6 - M^4 M_S^2 (x-1)^4 + M^2 (4x(2x-3)-1) M_S^4 (x-1)^2 + (4(3-2x)x+1) M_S^6 \big) \big] (l_T \cdot k_T)^2 \notag \\
& + l_T^2 \big[ 4 k_T^8 + \big( 12 M^2 (x-1)^2 + (18-5x) l_T^2 - 4 M_S^2 \big) k_T^6 + \big( 12 M^4 (x-1)^4 - 8 M^2 M_S^2 (x-1)^2 \notag \\
& + 2 (6-7x) l_T^4 - 4 M_S^4 + l_T^2 \big( (x(4x-7)-28) M_S^2 - M^2 (x-1)^2 (9x-28) \big) \big) k_T^4 + \big( 4 M^6 (x-1)^6 \notag \\
& - M^4 (3x-10) l_T^2 (x-1)^4 - 2 M^2 (5x-4) l_T^4 (x-1)^2 + (2-3x) l_T^6 + 4 M_S^6 \notag \\
& + \big( (x(40x-19)+2) l_T^2 - 4 M^2 (x-1)^2 \big) M_S^4 + 2 \big( -2 M^4 (x-1)^4 + M^2 (x-2)(4x+3) l_T^2 (x-1)^2 \notag \\
& + x(8x-7) l_T^4 \big) M_S^2 \big) k_T^2 + x l_T^2 \big( M^2 (x-1)^2 - M_S^2 \big) \big( M^2 (x-1)^2 + l_T^2 + 3 M_S^2 \big) \notag \\
&\times \big( M^2 (x-1)^2 - l_T^2 + (4x-5) M_S^2 \big) \big] l_T \cdot k_T + k_T^2 l_T^4 \big[ 4 k_T^6 + \big( 8 M^2 (x-1)^2 + (8-5x) l_T^2 \big) k_T^4 \notag \\
& + \big( 4 M^4 (x-1)^4 + (2-3x) l_T^4 - 4 M_S^4 + 2 l_T^2 \big( x(2x-5) M_S^2 - M^2 (x-1)^2 (3x-4) \big) \big) k_T^2 \notag \\
& - x l_T^2 \big( M^2 (x-1)^2 + l_T^2 + 3 M_S^2 \big) \big( M^2 (x-1)^2 + (3-4x) M_S^2 \big) \big] \bigg\}\left(256 M^4 \pi^3 (x-1) x k_T^2 M_S^4\right)^{-1}\,,
\end{align}

\begin{align}
&\mathcal{C}_{123}^{[f_{1T}^\perp]}+\mathcal{C}_{132}^{[f_{1T}^\perp]}+\mathcal{C}_{213}^{[f_{1T}^\perp]}+\mathcal{C}_{231}^{[f_{1T}^\perp]}+\mathcal{C}_{321}^{[f_{1T}^\perp]}+\mathcal{C}_{312}^{[f_{1T}^\perp]}\notag\\
=& \frac{3}{64 M^2 \pi^3 (x-1) x k_T^2 M_S^4} \times \bigg\{ 16 M_S^2 (l_T \cdot k_T)^4 \notag \\
& + 4 \Big( 2 ((2-3x) x+3) M_S^4 + \big(2 M^2 (x-1)^4+2 (x^2+3) k_T^2+((5-2x) x+2) l_T^2\big) M_S^2 \notag \\
& - (3x-2) \big(M^2 (x-1)^2+k_T^2\big) l_T^2 \Big) (l_T \cdot k_T)^3 - 2 \Big( \big(-M^2 (x-3) (2x-1) (x-1)^2 \notag \\
& +((17-4x) x-11) k_T^2+(x (6x-13)+3) M_S^2\big) l_T^4 + \Big(5 M^4 (x-1)^5+2 M^2 (9x-7) k_T^2 (x-1)^2 \notag \\
& +(13x-9) k_T^4+(x (22x-31)+3) M_S^4 -2 \big(M^2 (5 (x-1) x-1) (x-1)^2+(9 (x-1) x+7) k_T^2\big) M_S^2\Big) l_T^2 \notag \\
& + 2 M_S^2 \Big(M^4 (x-1)^5-4 M^2 k_T^2 (x-1)^2-(x+3) k_T^4+(x (4x-5)-1) M_S^4 \notag \\
& -2 \big(M^2 (2 (x-1) x-1) (x-1)^2+(x (2x-7)+8) k_T^2\big) M_S^2\Big) \Big) (l_T \cdot k_T)^2 \notag \\
& + \Big( (x-2) \big(M^2 x (x-1)^2+(7x-6) k_T^2-x M_S^2\big) l_T^6 + \Big(-M^4 x (x-1)^5 \notag \\
& +6 M^2 ((x-5) x+3) k_T^2 (x-1)^2+(x (7x-47)+34) k_T^4 + x (x (4x-15)+13) M_S^4 \notag \\
& -2 \big(M^2 x (2 (x-4) x+7) (x-1)^2+(x (2 (x-9) x+21)-9) k_T^2\big) M_S^2\Big) l_T^4 \notag \\
& + \Big( x (4 x (2x-5)+15) M_S^6 + \big((x (-8 (x-2) x-7)+10) k_T^2-M^2 (x-1)^2 x (8 (x-2) x+13)\big) M_S^4 \notag \\
& + \big( M^4 x (4x-3) (x-1)^4+2 M^2 (x (12x-7)-2) k_T^2 (x-1)^2+5 (x (4x-7)+4) k_T^4 \big) M_S^2 \notag \\
& + \big(M^2 (x-1)^2+k_T^2\big)^2 \big(M^2 x (x-1)^2+(10-13x) k_T^2\big) \Big) l_T^2 \notag \\
& + 2 M_S^2 \Big( (x+2) k_T^6 + \big(M^2 (x+4) (x-1)^2+(x (8x-21)+20) M_S^2\big) k_T^4 \notag \\
& + \big( -M^4 (x-2) (x-1)^4+2 M^2 (x (4x-5)-2) M_S^2 (x-1)^2+((11-8x) x+2) M_S^4 \big) k_T^2 \notag \\
& - x \big(M^2 (x-1)^2-M_S^2\big)^3 \Big) \Big) l_T \cdot k_T \notag \\
& + k_T^2 l_T^2 \Big( 2 x M_S^6 - 4 \big(M^2 x (x-1)^2+((7-4x) x-4) k_T^2\big) M_S^4 \notag \\
& + 2 x \big(M^2 (x-1)^2+k_T^2\big)^2 M_S^2 + l_T^4 \big( M^2 x (2x-3) (x-1)^2+(x (6x-19)+12) k_T^2 \notag \\
& +x (-4 (x-3) x-7) M_S^2 \big) + l_T^2 \big( -M^4 x (x-1)^4-2 M^2 (7x-6) k_T^2 (x-1)^2 \notag \\
& +4 \big(3 M^2 x (x-1)^2+(5x-6) k_T^2\big) M_S^2 (x-1) + (12-13x) k_T^4+x (4 (5-2x) x-11) M_S^4 \big) \Big) \bigg\}\,.
\end{align}

\subsection{Propeller function \texorpdfstring{$h_{1L}^\perp$}{h1L}}
When the coupling constant $\kappa_2$ is set to zero, or $ijk=\{111,333,113,131,311,133,313,331\}$, we can obtain $\mathcal{C}_{ijk}^{[h_{1L}^\perp]}$ by using $\mathcal{C}_{ijk}^{[f_{1T}^\perp]}$ in Eqs.~\eqref{f1T:111},~\eqref{f1T:333},~\eqref{f1T:113},~\eqref{f1T:133}
\begin{align}
\mathcal{C}_{ijk}^{[h_{1L}^\perp]}=\frac{4M^2(1-x) (k_T^2 l_T^2-(l_T \cdot k_T)^2)}{(k_T^2+l_T \cdot k_T)(k_T^2 (l_T^2+l_T \cdot k_T)+l_T \cdot k_T (M^2(1-x)^2-M_S^2))}\mathcal{C}_{ijk}^{[f_{1T}^\perp]}\,.
\end{align}
Then,
\begin{align}
&\mathcal{C}_{222}^{[h_{1L}^\perp]}\notag\\
=&3 M^2 (1-x) \bigg\{k_T^2 \Big(x l_T\cdot k_T \left(l_T^2\left((4-5 x) M_S^2-2 M^2 (x-1)^2\right)+2 x M_S^2 \left(M_S^2-M^2(x-1)^2\right)\right)\notag\\
&+2 \left(l_T\cdot k_T\right){}^2 \left((1-2x) l_T^2-2 (x-1) x M_S^2\right)+x l_T^2 \left(M^2(x-1)^2-M_S^2\right) \left(l_T^2+x M_S^2\right)\Big)+2 x\left(l_T\cdot k_T\right){}^2\notag\\
&\times \left(x M_S^2 \left(2 l_T\cdot k_T-M^2 (x-1)^2+M_S^2\right)+l_T^2 \left(M_S^2-M^2(x-1)^2\right)\right)\notag\\
&+k_T^4 \left(-2 (x-2) x M_S^2 l_T\cdot k_T+x(x+2) l_T^2 M_S^2+(4 x-2) l_T^4\right)\bigg\}\left(8 \pi ^3 x k_T^4M_S^4\right)^{-1}\,,
\end{align}

\begin{align}
&\mathcal{C}_{112}^{[h_{1L}^\perp]}+\mathcal{C}_{121}^{[h_{1L}^\perp]}+\mathcal{C}_{211}^{[h_{1L}^\perp]}\notag\\
=&3 M^2 (x-1) \bigg\{x k_T^2 l_T^4 \left(M_S^2-M^2(x-1)^2\right)+2 x k_T^2 l_T\cdot k_T \left(M^2 (x-1)^2 l_T^2-2M_S^2 \left(k_T^2+(1-2 x) l_T^2\right)\right)\notag\\
&+2 \left(l_T\cdot k_T\right){}^2 \left(l_T^2 \left((5 x-4) k_T^2+M^2 x (x-1)^2-xM_S^2\right)+2 (x-2) k_T^2 M_S^2\right)\notag\\
&-8 x^2 M_S^2 \left(l_T\cdot k_T\right){}^3-2 (5 x-4) k_T^4 l_T^2\left(l_T^2+M_S^2\right)\bigg\}\left(4 \pi ^3 x k_T^4 M_S^4\right)^{-1}\,,
\end{align}

\begin{align}
&\mathcal{C}_{223}^{[h_{1L}^\perp]}+\mathcal{C}_{232}^{[h_{1L}^\perp]}+\mathcal{C}_{322}^{[h_{1L}^\perp]}\notag\\
=& -3 \bigg\{ k_T^2 l_T^4 \Big( 2 k_T^2 \big( (-2 (x-3) x-3) l_T^2 + (x-1) (2 M^2 (x-1) (3 x-2) - x (x+4) M_S^2) \big)+ 2 (5 x-4) k_T^4 \notag \\
&  + x (M^2 (x-1)^2 - M_S^2) \big( -(x-3) l_T^2 + 2 M^2 (x-1)^2 + 2 (-x^2+x+1) M_S^2 \big) \Big)+ 8 (l_T \cdot k_T)^3 \Big( x M_S^2 (2 x k_T^2  \notag \\
&+ (1-(x-2) x) l_T^2  + M^2 (x-1)^3) + l_T^2 ((1-2 x) k_T^2 - M^2 (x-1)^2 x) - (x-1) x (2 x^2+1) M_S^4 \Big) \notag \\
& + 2 l_T^2 (l_T \cdot k_T)^2 \Big( k_T^2 ((x-3) (2 x-1) l_T^2 - 4 M^2 (3 x-1) (x-1)^2 + 2 x (x (2 x-7)+7) M_S^2) + (4-10 x) k_T^4 \notag \\
& - x (M^2 (x-1)^2 - M_S^2) \big( -(x-3) l_T^2 + 2 M^2 (x-1)^2 + 2 (-x^2+x+1) M_S^2 \big) \Big) \notag \\
& - 2 k_T^2 l_T^2 l_T \cdot k_T \Big( 2 (x-1) k_T^2 (-2 l_T^2 + 2 M^2 (x-1) x - (x-6) x M_S^2) + 2 x k_T^4 \notag \\
& + x \big( -M_S^2 ((x (5 x-12)+5) l_T^2 + 2 M^2 (x-2) (x-1)^3) + M^2 (x-1)^3 (2 M^2 (x-1) - l_T^2)  \notag \\
& + 2 (x ((5-4 x) x-3)+1) M_S^4 \big) \Big)+ 16 x^2 M_S^2 (l_T \cdot k_T)^4 \bigg\}\left(32 \pi^3 x k_T^4 M_S^4\right)^{-1}\,,
\end{align}

\begin{align}
&\mathcal{C}_{221}^{[h_{1L}^\perp]}+\mathcal{C}_{212}^{[h_{1L}^\perp]}+\mathcal{C}_{122}^{[h_{1L}^\perp]}\notag\\
=& 3 M^2 (1-x) \Bigg\{ k_T^2 l_T^2 \bigg( l_T^2 \Big( 2 (5-8 x) k_T^2 + 3 x \big( M_S^2 - M^2 (x-1)^2 \big) \Big) - 2 M_S^2 \Big( (x (x+5)-2) k_T^2 + x^2 \big( M^2 (x-1)^2 - M_S^2 \big) \Big) \bigg) \notag \\
& + 2 (l_T \cdot k_T)^2 \bigg( l_T^2 \Big( (8 x-5) k_T^2 + 3 M^2 x (x-1)^2 - 3 x M_S^2 \Big) - 2 M_S^2 \Big( (1-2 (x-1) x) k_T^2 + x^2 \big( M_S^2 - M^2 (x-1)^2 \big) \Big) \bigg) \notag \\
& + 2 x k_T^2 l_T \cdot k_T \bigg( 2 M_S^2 \Big( (x-3) k_T^2 + M^2 x (x-1)^2 - x M_S^2 \Big)  + 3 l_T^2 \Big( M^2 (x-1)^2 + (3 x-2) M_S^2 \Big) \bigg)\notag\\
&  - 16 x^2 M_S^2 (l_T \cdot k_T)^3 \Bigg\}\left(8 \pi^3 x k_T^4 M_S^4\right)^{-1}\,,
\end{align}

\begin{align}
&\mathcal{C}_{233}^{[h_{1L}^\perp]}+\mathcal{C}_{323}^{[h_{1L}^\perp]}+\mathcal{C}_{332}^{[h_{1L}^\perp]}\notag\\
=& -3 \Bigg\{ 4 (l_T \cdot k_T)^3 \bigg( -l_T^2 M_S^2 \Big( 4 (x^2-1) k_T^2 + x (2 x+1) l_T^2 + 2 M^2 x (x-1)^3 \Big) \notag \\
& + l_T^4 \big( (2 x-1) k_T^2 + M^2 x (x-1)^2 \big) - 2 k_T^2 l_T^2 \big( k_T^2 + M^2 (x-1)^2 \big) - 2 x M_S^4 \big( 4 x k_T^2 + (3 x+1) l_T^2 \big) \bigg) \notag \\
& + k_T^2 l_T^4 \bigg( k_T^4 \big( (8-5 x) l_T^2 + 8 M^2 (x-1)^2 \big) + k_T^2 \Big( 2 l_T^2 \big( x (2 x-5) M_S^2 - M^2 (x-1)^2 (3 x-4) \big) \notag \\
& + (2-3 x) l_T^4 + 4 M^4 (x-1)^4 + 4 (4 (x-1) x-1) M_S^4 \Big) + 4 k_T^6 - x l_T^2 \big( M^2 (x-1)^2 - M_S^2 \big) \big( l_T^2 + M^2 (x-1)^2 + 3 M_S^2 \big) \bigg) \notag \\
& + 2 l_T^2 (l_T \cdot k_T)^2 \bigg( -k_T^4 \big( (4-5 x) l_T^2 + 4 M^2 (x-1)^2 + 8 M_S^2 \big) + k_T^2 \Big( 2 (3 x-2) l_T^2 \big( M^2 (x-1)^2 + x M_S^2 \big) \notag \\
& + (2 x-1) l_T^4 - 2 M^4 (x-1)^4 + 2 (4 (x-2) x+1) M_S^4 \Big) - 2 k_T^6 + x l_T^2 \big( M^2 (x-1)^2 - M_S^2 \big) \big( l_T^2 + M^2 (x-1)^2 + 3 M_S^2 \big) \bigg) \notag \\
& + 2 k_T^2 l_T^2 l_T \cdot k_T \bigg( 2 l_T^2 M_S^2 \Big( (x (4 x-3)-4) k_T^2 + x (2 x-1) \big( l_T^2 + M^2 (x-1)^2 \big) \Big) \notag \\
& + l_T^2 \Big( 2 (x-1) k_T^2 \big( M^2 (x-1) (x+2) - l_T^2 \big) + (x+4) k_T^4 + M^4 x (x-1)^4 \Big) + x M_S^4 \big( 16 (x-1) k_T^2 + (12 x-7) l_T^2 \big) \bigg) \notag \\
& - 16 M_S^2 (l_T \cdot k_T)^4 \big( x^2 (l_T^2 + 2 M_S^2) - k_T^2 \big) \Bigg\}\left(64 \pi^3 M^2 x k_T^4 M_S^4\right)^{-1}\,,
\end{align}

\begin{align}
&\mathcal{C}_{123}^{[h_{1L}^\perp]}+\mathcal{C}_{132}^{[h_{1L}^\perp]}+\mathcal{C}_{213}^{[h_{1L}^\perp]}+\mathcal{C}_{231}^{[h_{1L}^\perp]}+\mathcal{C}_{321}^{[h_{1L}^\perp]}+\mathcal{C}_{312}^{[h_{1L}^\perp]}\notag\\
=& 3 \Bigg\{ 4 (l_T \cdot k_T)^3 \bigg( l_T^2 \big( (3-4 x) k_T^2 - M^2 x (x-1)^2 + x (1-2 (x-2) x) M_S^2 \big)  + 2 x M_S^2 \big( (2 x+1) k_T^2 + M^2 x (x-1)^2\notag\\
& + (3-2 x) x M_S^2 \big) \bigg) - 2 k_T^2 l_T \cdot k_T \bigg( l_T^4 \big( -6 (x-1) k_T^2 - M^2 x (x-1)^3 - 2 (x-2) x (2 x-1) M_S^2 \big) \notag \\
& + x l_T^2 \Big( 2 M_S^2 \big( (5 x-4) k_T^2 + M^2 x (x-1)^2 \big) + \big( k_T^2 + M^2 (x-1)^2 \big)^2 + (2 (7-4 x) x-5) M_S^4 \Big) \notag \\
& - 2 x M_S^2 \big( k_T^2 + M^2 (x-1)^2 \big)^2 + 4 x M_S^4 \big( M (x-1) - k_T \big) \big( k_T + M (x-1) \big) - 2 x M_S^6 \bigg) - 2 (l_T \cdot k_T)^2 \notag\\
&\times \bigg( l_T^4 \big( (-3 (x-3) x-5) k_T^2 + (x-2) x \big( M_S^2 - M^2 (x-1)^2 \big) \big)  + l_T^2 \Big( 2 k_T^2 \big( M^2 (5 x-3) (x-1)^2 \notag\\
&+ (x^2+x-3) M_S^2 \big) + (9 x-6) k_T^4  + x \big( M^4 (x-1)^4 - 2 M^2 (x-1)^3 M_S^2 + (2 x-3) M_S^4 \big) \Big) \notag \\
& - 2 x M_S^2 \Big( 4 k_T^2 \big( M^2 (x-1)^2 + x M_S^2 \big) + 3 k_T^4 + \big( M_S^2 - M^2 (x-1)^2 \big)^2 \Big) \bigg) \notag \\
& + k_T^2 l_T^2 \bigg( - (x-2) l_T^4 \big( (6 x-5) k_T^2 + M^2 x (x-1)^2 - x M_S^2 \big)  + l_T^2 \Big( 2 (x-1) k_T^2 \big( M^2 (x-1) (7 x-6) + (6-9 x) M_S^2 \big)\notag\\
& + (13 x-12) k_T^4  + x \big( M^4 (x-1)^4 - 2 M^2 (x-1)^3 M_S^2 + (2 x-3) M_S^4 \big) \Big)  - 2 x M_S^2 \Big( 2 k_T^2 \big( M^2 (x-1)^2 + (4 x-3) M_S^2 \big)\notag\\
& + k_T^4 + \big( M_S^2 - M^2 (x-1)^2 \big)^2 \Big) \bigg)  + 16 x^2 M_S^2 (l_T \cdot k_T)^4 \Bigg\}\left(16 \pi^3 x k_T^4 M_S^4\right)^{-1}\,,
\end{align}

\subsection{Linearity function \texorpdfstring{$h_1$}{h1}}
When the coupling constant $\kappa_2$ is set to zero, or $ijk=\{111,333,113,131,311,133,313,331\}$, we can obtain $\mathcal{C}_{ijk}^{[h_1]}$ by using $\mathcal{C}_{ijk}^{[f_{1T}^\perp]}$ in Eqs.~\eqref{f1T:111},~\eqref{f1T:333},~\eqref{f1T:113},~\eqref{f1T:133} and the relation of Eq.~\eqref{eq:g2}
\begin{align}
\mathcal{C}_{ijk}^{[h_1]}=5~\mathcal{C}_{ijk}^{[f_{1T}^\perp]}\,.
\end{align}
Then,
\begin{align}
&\mathcal{C}_{222}^{[h_1]}\notag\\
=& -3 \Bigg\{ k_T^4 l_T^2 \bigg( M_S^2 \Big( (x (3 x-7)+10) k_T^2 + 3 x \big( 2 x l_T^2 + M^2 (x-3) (x-1)^2 \big) \Big) \notag \\
& - 2 (x-1) l_T^2 \big( 5 k_T^2 + 3 M^2 (x-1) x \big) - 3 (x-3) x M_S^4 \bigg) \notag \\
& + k_T^2 (l_T \cdot k_T)^2 \bigg( 2 l_T^2 \Big( (5-11 x) k_T^2 - 6 M^2 x (x-1)^2 + 2 x (x+5) M_S^2 \Big) \notag \\
& + M_S^2 \Big( (x (5 x-29)+30) k_T^2 + x (7 x-13) \big( M^2 (x-1)^2 - M_S^2 \big) \Big) \bigg) \notag \\
& + k_T^2 l_T \cdot k_T \bigg( 2 k_T^2 \Big( l_T^2 \big( (x (4 x-3)+5) M_S^2 - 9 M^2 (x-1)^2 x \big) + (5-8 x) l_T^4 \notag \\
& + x (5 x-11) M_S^2 \big( M^2 (x-1)^2 - M_S^2 \big) \Big) + 2 k_T^4 \big( (5-8 x) l_T^2 + (x-2) (2 x-5) M_S^2 \big) \notag \\
& + 3 x \big( M^2 (x-1)^2 - M_S^2 \big) \Big( M_S^2 \big( (x-3) l_T^2 + 2 M^2 x (x-1)^2 \big) - 2 l_T^4 - 2 x M_S^4 \Big) \bigg) \notag \\
& + 2 M_S^2 (l_T \cdot k_T)^3 \Big( (x (3 x+4)+5) k_T^2 + M^2 x (x-1)^3 - x (x-1) M_S^2 \Big) \Bigg\}\left(16 \pi^3 x k_T^4 M_S^4\right)^{-1}\,,
\end{align}

\begin{align}
&\mathcal{C}_{112}^{[h_1]}+\mathcal{C}_{121}^{[h_1]}+\mathcal{C}_{211}^{[h_1]}\notag\\
=& 3 \Bigg\{ (x-1) k_T^4 l_T^2 \bigg( l_T^2 \Big( 25 k_T^2 + 6 M^2 (x-1) x - 6 x M_S^2 \Big) + 6 M_S^2 \Big( 5 k_T^2 + 2 M^2 (x-1) x - 2 x M_S^2 \Big) \bigg) \notag \\
& + k_T^2 (l_T \cdot k_T)^2 \bigg( l_T^2 \Big( (31x-25) k_T^2 + 3 M^2 (7x-5) (x-1)^2 + (15 - x(4x+29)) M_S^2 \Big) \notag \\
& + 2 M_S^2 \Big( 3 (3x-5) k_T^2 + M^2 (11x-5) (x-1)^2 + (x(10x-21)+5) M_S^2 \Big) \bigg) \notag \\
& - 4 x M_S^2 (l_T \cdot k_T)^3 \Big( (x+5) k_T^2 + M^2 (x-1)^3 - (x-1) M_S^2 \Big) \notag \\
& + k_T^2 l_T \cdot k_T \bigg( l_T^4 \Big( (28x-25) k_T^2 + 3 M^2 x (x-1)^2 - 3 x M_S^2 \Big) \notag \\
& + l_T^2 \Big( 3 (x-1) k_T^2 \big( M^2 (x-1) (9x-5) + (5-2x) M_S^2 \big) + (28x-25) k_T^4 \notag \\
& + 3 x \big( -M^4 (x-1)^4 - 2 M^2 (x-2) (x-1)^2 M_S^2 + (2x-3) M_S^4 \big) \Big) \notag \\
& + 2 M_S^2 \Big( k_T^2 \big( M^2 (17x-5) (x-1)^2 + (x(4x-15)+5) M_S^2 \big) + 3 (4x-5) k_T^4 \notag \\
& + 3 x \big( M^4 (x-1)^4 - 2 M^2 (x-1)^3 M_S^2 + (2x-3) M_S^4 \big) \Big) \bigg) \Bigg\}\left(8\pi^3 x k_T^4 M_S^4\right)^{-1}\,,
\end{align}

\begin{align}
&\mathcal{C}_{223}^{[h_1]}+\mathcal{C}_{232}^{[h_1]}+\mathcal{C}_{322}^{[h_1]}\notag\\
=& 3 \bigg\{ 4 M_S^2 \big( M^2 x (x-1)^3 - x M_S^2 (x-1) + (x^2+5) k_T^2 \big) (l_T \cdot k_T)^4 \notag \\
& + 2 \Big( 2 \big( (3 x^2 + x + 5) M_S^2 - 3 x l_T^2 \big) k_T^4 + \big( -x ( x (2 x (x+1) + 21) - 31 ) M_S^4 + M^2 (x-1)^2 x (7 x - 13) M_S^2 \notag \\
& + l_T^2 \big( ( x ((2-3 x) x + 9) + 10 ) M_S^2 - 6 M^2 (x-1)^2 x \big) \big) k_T^2 + (x-1) x M_S^2 \big( M^2 (x-1)^2 - M_S^2 \big) \notag \\
& \big( M^2 (x-1)^2 - (x-2) l_T^2 + (-2 (x-1) x - 1) M_S^2 \big) \Big) (l_T \cdot k_T)^3 \notag \\
& - k_T^2 l_T^2 \big( 5 M^4 (x-1)^5 + M^2 ( x (x (7 x - 33) + 34) + 10 ) M_S^2 (x-1)^2 + (29 x - 5) k_T^4 \notag \\
& + ( x (x (x (4 x + 5) + 33) - 55) - 5 ) M_S^4 + 2 x l_T^2 \big( (2 x (x+2) - 15) M_S^2 - 3 M^2 (x-3) (x-1)^2 \big) \notag \\
& + k_T^2 \big( 2 M^2 (17 x - 5) (x-1)^2 - (x-3) (11 x - 5) l_T^2 + ( x (x (5 x - 57) + 40) - 30 ) M_S^2 \big) \big) (l_T \cdot k_T)^2 \notag \\
& + k_T^2 l_T^2 \big( (5 - 14 x) k_T^6 + ( -5 M^2 (5 x - 2) (x-1)^2 + (x (8 x - 51) + 25) l_T^2 \notag \\
& + ( 10 - (x-2) x (4 x - 17) ) M_S^2 ) k_T^4 + ( -M^4 (8 x - 5) (x-1)^4 - 2 M^2 ( (x-3) x (5 x - 4) + 5 ) M_S^2 (x-1)^2 \notag \\
& + (x-2) (8 x - 5) l_T^4 + ( 2 (-3 x^2 + x + 4) x^2 + 5 ) M_S^4 + l_T^2 ( M^2 ( x (9 x - 37) + 10 ) (x-1)^2 \notag \\
& + ( x ((37 - 8 x) x - 21) + 10 ) M_S^2 ) ) k_T^2 + 3 x ( M^2 (x-1)^2 - M_S^2 ) ( M^4 (x-1)^4 \notag \\
& - M^2 (2 x - 1) M_S^2 (x-1)^3 + (x-2) l_T^4 + x (-2 (x-1) x - 1) M_S^4 - (x-3) (x-2) l_T^2 M_S^2 ) \big) l_T \cdot k_T \notag \\
& + k_T^4 l_T^4 \big( (10 - 13 x) k_T^4 + ( (x (5 x - 18) + 10) l_T^2 + (x-1) ( ((10 - 3 x) x - 10) M_S^2 \notag \\
& - 2 M^2 (x-1) (8 x - 5) ) ) k_T^2 + 3 x ( -M^4 (x-1)^4 + M^2 (x-2) l_T^2 (x-1)^2 + (-2 x^3 + x^2 + x + 1) M_S^4 \notag \\
& - ( M^2 (x-4) (x-1)^3 + (2 (x-2) x + 1) l_T^2 ) M_S^2 ) \big) \bigg\}\left(32 M^2 \pi^3 (1-x) x k_T^4 M_S^4\right)^{-1}\,,
\end{align}

\begin{align}
&\mathcal{C}_{221}^{[h_1]}+\mathcal{C}_{212}^{[h_1]}+\mathcal{C}_{122}^{[h_1]}\notag\\
=& 3 \bigg\{ k_T^4 l_T^2 \bigg( l_T^2 \big( -20 (x-1) k_T^2 - 9 M^2 x (x-1)^2 + 3 x (3x-2) M_S^2 \big) \notag \\
& + M_S^2 \big( (x-4)(3x-5) k_T^2 + 3 M^2 (x-5) x (x-1)^2 + x (x(8x-5)+3) M_S^2 \big) \bigg) \notag \\
& + k_T^2 (l_T \cdot k_T)^2 \bigg( l_T^2 \big( 4 (5-8x) k_T^2 - M^2 (17x-5) (x-1)^2 + (x+5)(6x-1) M_S^2 \big) \notag \\
& + M_S^2 \big( (x(5x-27)+40) k_T^2 + M^2 x (7x-13) (x-1)^2 + (23-17x) x M_S^2 \big) \bigg) \notag \\
& + 2 M_S^2 (l_T \cdot k_T)^3 \big( (x+1)(4x+5) k_T^2 + 2 M^2 x (x-1)^3 - 2 x (x-1) M_S^2 \big) \notag \\
& + k_T^2 l_T \cdot k_T \bigg( l_T^4 \big( (20-26x) k_T^2 + 6 x (M_S^2 - M^2 (x-1)^2) \big) \notag \\
& + l_T^2 \Big( k_T^2 \big( (x(11x-17)+15) M_S^2 - M^2 (x-1)^2 (26x-5) \big) + (20-26x) k_T^4 \notag \\
& + 3 x \big( M^4 (x-1)^4 + 2 M^2 (x-2) (x-1)^2 M_S^2 + (3-2x) M_S^4 \big) \Big) \notag \\
& + 2 M_S^2 \Big( x k_T^2 \big( M^2 (5x-14) (x-1)^2 + (x(4x-11)+13) M_S^2 \big) + (x-3)(2x-5) k_T^4 \notag \\
& + (x-1) x^2 \big( 3 M^4 (x-1)^3 + M^2 (x-1)(4x-7) M_S^2 - 4 M_S^4 \big) \Big) \bigg) \bigg\}\left(8\pi^3 x k_T^4 M_S^4\right)^{-1}\,,
\end{align}

\begin{align}
&\mathcal{C}_{233}^{[h_1]}+\mathcal{C}_{323}^{[h_1]}+\mathcal{C}_{332}^{[h_1]}\notag\\
=& 3 \bigg\{ 8 M_S^2 \big( -5 k_T^4 + ( -5 M^2 (x-1)^2 + (x^2+5) l_T^2 + (2 x^2+15) M_S^2 ) k_T^2 \notag \\
& + (x-1) x (M^2 (x-1)^2 - M_S^2) (l_T^2 + 2 M_S^2) \big) (l_T \cdot k_T)^4 + 4 \big( 5 (l_T^2 - 2 M_S^2) k_T^6 \notag \\
& + ( -3 x l_T^4 + ( 10 M^2 (x-1)^2 + (6 x^2+2 x-15) M_S^2 ) l_T^2 + (4 x (5 x-1)+30) M_S^4 - 10 M^2 (x-1)^2 M_S^2 ) k_T^4 \notag \\
& + ( ( (x (3 x+7)+5) M_S^2 - 3 M^2 (x-1)^2 x ) l_T^4 + ( 5 M^4 (x-1)^4 + M^2 (x (7 x-13)-15) M_S^2 (x-1)^2 \notag \\
& + (x (x+21)+30) M_S^4 ) l_T^2 + 4 x (5 x-8) M_S^4 (M^2 (x-1)^2 - M_S^2) ) k_T^2 \notag \\
& + (x-1) x l_T^2 M_S^2 (M^2 (x-1)^2 - M_S^2) (M^2 (x-1)^2 + l_T^2 + 3 M_S^2) \big) (l_T \cdot k_T)^3 \notag \\
& + k_T^2 l_T^2 \big( 10 M^6 (x-1)^6 + 30 k_T^6 + k_T^4 ( 70 M^2 (x-1)^2 + (35-29 x) l_T^2 - 110 M_S^2 ) \notag \\
& + 4 x l_T^4 ( (x+5) M_S^2 - 3 M^2 (x-1)^2 ) + l_T^2 ( -5 M^4 (x-3) (x-1)^4 + 2 M^2 (13 (x-2) x-15) M_S^2 (x-1)^2 \notag \\
& + ((89-10 x) x+15) M_S^4 ) + 2 M_S^2 ( (8 x (4 x-7)-5) M_S^2 (M^2 (x-1)^2 - M_S^2) - 5 M^4 (x-1)^4 ) \notag \\
& + 2 k_T^2 ( 25 M^4 (x-1)^4 - 40 M^2 M_S^2 (x-1)^2 + l_T^2 ( (23 x+25) M_S^2 - M^2 (x-1) (17 x-25) ) (x-1) \notag \\
& + (5-11 x) l_T^4 + (8 x (9 x-2)+55) M_S^4 ) \big) (l_T \cdot k_T)^2  + k_T^2 l_T^2 \big( 10 k_T^8 + ( 30 M^2 (x-1)^2 + (45-14 x) l_T^2 - 10 M_S^2 ) k_T^6 \notag \\
& + ( 30 M^4 (x-1)^4 - 20 M^2 M_S^2 (x-1)^2 + (30-38 x) l_T^4 - 10 M_S^4 \notag \\
& + l_T^2 ( (x (6 x-17)-70) M_S^2 - 5 M^2 (x-1)^2 (5 x-14) ) ) k_T^4  + ( (5-8 x) l_T^6 - 4 (7 x-5) (M^2 (x-1)^2 - x M_S^2) l_T^4 \notag \\
& + ( -M^4 (8 x-25) (x-1)^4 + 2 M^2 (2 x (3 x-5)-15) M_S^2 (x-1)^2 + (4 x (19 x-3)+5) M_S^4 ) l_T^2 \notag \\
& + 10 (M_S^2 - M^2 (x-1)^2)^2 (M^2 (x-1)^2 + M_S^2) ) k_T^2 \notag \\
& + 3 x l_T^2 (M^2 (x-1)^2 - M_S^2) (M^2 (x-1)^2 + l_T^2 + 3 M_S^2) (M^2 (x-1)^2 - l_T^2 + (2 x-3) M_S^2) \big) l_T \cdot k_T \notag \\
& + k_T^4 l_T^4 \big( 10 k_T^6 + ( 20 M^2 (x-1)^2 + (20-13 x) l_T^2 ) k_T^4 \notag \\
& + ( 10 M^4 (x-1)^4 + (5-8 x) l_T^4 - 10 M_S^4 + 2 l_T^2 ( x (3 x-11) M_S^2 - 2 M^2 (x-1)^2 (4 x-5) ) ) k_T^2 \notag \\
& - 3 x l_T^2 (M^2 (x-1)^2 + l_T^2 + 3 M_S^2) (M^2 (x-1)^2 + (1-2 x) M_S^2) \big) \bigg\}\left(128 M^4 \pi^3 (x-1) x k_T^4 M_S^4\right)^{-1}\,,
\end{align}

\begin{align}
&\mathcal{C}_{123}^{[h_1]}+\mathcal{C}_{132}^{[h_1]}+\mathcal{C}_{213}^{[h_1]}+\mathcal{C}_{231}^{[h_1]}+\mathcal{C}_{321}^{[h_1]}+\mathcal{C}_{312}^{[h_1]}\notag\\
=& 3 \bigg\{ 8 M_S^2 \Big( M^2 x (x-1)^3 - x M_S^2 (x-1) + (x^2+5) k_T^2 \Big) (l_T \cdot k_T)^4 \notag \\
& + 4 \Big( \big( (5-8x) l_T^2 + 3 (2 x^2+x+5) M_S^2 \big) k_T^4 + \big( (x (12-x (2x+13)) + 15) M_S^4 \notag \\
& + M^2 (x-1)^3 (7x-5) M_S^2 - l_T^2 \big( M^2 (8x-5) (x-1)^2 + ((x-3) x (x+4)-5) M_S^2 \big) \Big) k_T^2 \notag \\
& + (x-1) x M_S^2 \big( M^2 (x-1)^2 - M_S^2 \big) \big( M^2 (x-1)^2 - (x-2) l_T^2 + (3-2x) M_S^2 \big) \Big) (l_T \cdot k_T)^3 \notag \\
& + k_T^2 \Big( \big( M^2 (x-3) (11x-5) (x-1)^2 + (x (21x-88)+55) k_T^2 - (x (x (4x+21)-64)+15) M_S^2 \big) l_T^4 \notag \\
& + \big( -25 M^4 (x-1)^5 - 2 M^2 (47x-35) k_T^2 (x-1)^2 + (45-69x) k_T^4 \notag \\
& - \big( x (8x^2+90x-149) + 15 \big) M_S^4 + 2 \big( M^2 (x (25x-26)-5) (x-1)^2 + (x (43x-36)+35) k_T^2 \big) M_S^2 \big) l_T^2 \notag \\
& + 2 M_S^2 \big( -5 M^4 (x-1)^5 + 2 M^2 (x (10x-11)-5) M_S^2 (x-1)^2 + 3 (3x+5) k_T^4 \notag \\
& + ((27-20x)x+5) M_S^4 + 4 k_T^2 \big( M^2 (x+5) (x-1)^2 + (x (5x-16)+20) M_S^2 \big) \Big) \Big) (l_T \cdot k_T)^2 \notag \\
& + k_T^2 \Big( 6 x M_S^8 + 2 \big( 5 (-4x^2+6x+1) k_T^2 - 9 M^2 (x-1)^2 x \big) M_S^6 \notag \\
& + 2 \big( 9 M^4 x (x-1)^4 + 2 M^2 (2x (5x-7)-5) k_T^2 (x-1)^2 + (x (20x-49)+50) k_T^4 \big) M_S^4 \notag \\
& - 2 \big( M^2 (x-1)^2 + k_T^2 \big)^2 \big( 3 M^2 (x-1)^2 x - (4x+5) k_T^2 \big) M_S^2 \notag \\
& + 3 (x-2) l_T^6 \big( M^2 x (x-1)^2 + (6x-5) k_T^2 - x M_S^2 \big) \notag \\
& + l_T^2 \big( 3 x (2x (2x-5)+9) M_S^6 + \big( (25-4x (3 (x-2)x+1)) k_T^2 - 3 M^2 (x-1)^2 x (4 (x-2)x+9) \big) M_S^4 \notag \\
& + \big( 3 M^4 x (2x-1) (x-1)^4 + 2 M^2 (2x (15x-8)-5) k_T^2 (x-1)^2 + (x (46x-81)+50) k_T^4 \big) M_S^2 \notag \\
& + \big( M^2 (x-1)^2 + k_T^2 \big)^2 \big( 3 M^2 x (x-1)^2 + (25-34x) k_T^2 \big) \big) \notag \\
& + l_T^4 \big( (x (18x-121)+85) k_T^4 + \big( M^2 (x (17x-80)+45) (x-1)^2 + (x ((71-6x)x-86)+45) M_S^2 \big) k_T^2 \notag \\
& + 3 x \big( -M^4 (x-1)^5 - 2 M^2 ((x-5)x+5) M_S^2 (x-1)^2 + (x-3) (2x-3) M_S^4 \big) \big) \Big) l_T \cdot k_T \notag \\
& + k_T^4 l_T^2 \Big( 6 x M_S^6 + 4 \big( (x (10x-17)+10) k_T^2 - 3 M^2 (x-1)^2 x \big) M_S^4 \notag \\
& + 6 x \big( M^2 (x-1)^2 + k_T^2 \big)^2 M_S^2 + 3 l_T^4 \big( M^2 x (2x-3) (x-1)^2 + (x (5x-16)+10) k_T^2 \notag \\
& + x (-2 (x-3)x-3) M_S^2 \big) + l_T^2 \big( -3 M^4 x (x-1)^4 + 30 M^2 x M_S^2 (x-1)^3 \notag \\
& + 2 k_T^2 \big( (23x-30) M_S^2 - 3 M^2 (x-1) (6x-5) \big) (x-1) + (30-33x) k_T^4 \notag \\
& - 3 x (2x (2x-5)+5) M_S^4 \big) \Big) \bigg\}\left(32 M^2 \pi^3 (x-1) x k_T^4 M_S^4\right)^{-1}\,,
\end{align}

\subsection{Butterfly function \texorpdfstring{$h_{1T}^\perp$}{h1T}}
When the coupling constant $\kappa_2$ is set to zero, or $ijk=\{111,333,113,131,311,133,313,331\}$, we can obtain $\mathcal{C}_{ijk}^{[h_{1T}^\perp]}$ by using $\mathcal{C}_{ijk}^{[f_{1T}^\perp]}$ in Eqs.~\eqref{f1T:111},~\eqref{f1T:333},~\eqref{f1T:113},~\eqref{f1T:133} and the relation of Eq.~\eqref{eq:g2}
\begin{align}
\mathcal{C}_{ijk}^{[h_{1T}^\perp]}=-\frac{2M^2}{\bm{k}_T^2}~\mathcal{C}_{ijk}^{[f_{1T}^\perp]}\,.
\end{align}
Then,
\begin{align}
&\mathcal{C}_{222}^{[h_{1T}^\perp]}\notag\\
=& 3 M^2 (1-x) \Bigg\{ k_T^4 l_T^2 \Big( k_T^2 \big( 2 l_T^2 + (x+2) M_S^2 \big) + x M_S^2 \big( 2 l_T^2 + M^2 (x-1)^2 - M_S^2 \big) \Big) \notag \\
& + k_T^2 (l_T \cdot k_T)^2 \Big( k_T^2 \big( 2 l_T^2 + (6-5x) M_S^2 \big) + x M_S^2 \big( -4 l_T^2 - 3 M^2 (x-1)^2 + 3 M_S^2 \big) \Big) \notag \\
& + k_T^2 l_T \cdot k_T \Big( 2 k_T^2 \big( M_S^2 ((x+1) l_T^2 - M^2 (x-1)^2 x) + l_T^4 + x M_S^4 \big) \notag \\
& \quad + 2 k_T^4 \big( l_T^2 - (x-2) M_S^2 \big) + x l_T^2 M_S^2 \big( M^2 (x-1)^2 - M_S^2 \big) \Big) \notag \\
& + 2 M_S^2 (l_T \cdot k_T)^3 \big( (1-3x) k_T^2 + x (M_S^2 - M^2 (x-1)^2) \big) \Bigg\}\left(8 \pi^3 x k_T^6 M_S^4\right)^{-1}\,,
\end{align}

\begin{align}
&\mathcal{C}_{112}^{[h_{1T}^\perp]}+\mathcal{C}_{121}^{[h_{1T}^\perp]}+\mathcal{C}_{211}^{[h_{1T}^\perp]}\notag\\
=& 3 M^2 (1-x) \bigg\{ k_T^2 (l_T \cdot k_T)^2 \bigg( l_T^2 \big( 5 k_T^2 + 3 M^2 (x-1)^2 - (4x+3) M_S^2 \big) \notag \\
& + 2 M_S^2 \big( 3 k_T^2 + M^2 (x-1)^2 + (2x-1) M_S^2 \big) \bigg) - 4 x M_S^2 (l_T \cdot k_T)^3 \big( k_T^2 + M^2 (x-1)^2 - M_S^2 \big) \notag \\
& + l_T \cdot k_T \bigg( k_T^4 \Big( l_T^2 \big( 3 M^2 (x-1)^2 + (2x+3) M_S^2 \big) + 5 l_T^4 + 2 M^2 (x-1)^2 M_S^2 + (4x-2) M_S^4 \Big) \notag \\
& + 2 x k_T^2 l_T^2 M_S^2 \big( M^2 (x-1)^2 - M_S^2 \big) + k_T^6 \big( 5 l_T^2 + 6 M_S^2 \big) \bigg)  + 2 x k_T^4 l_T^4 M_S^2 + k_T^6 \big( 6 l_T^2 M_S^2 + 5 l_T^4 \big) \bigg\}\left(4 \pi^3 x k_T^6 M_S^4\right)^{-1}\,,
\end{align}

\begin{align}
&\mathcal{C}_{223}^{[h_{1T}^\perp]}+\mathcal{C}_{232}^{[h_{1T}^\perp]}+\mathcal{C}_{322}^{[h_{1T}^\perp]}\notag\\
=& -3 \bigg\{ -2 M_S^2 (l_T \cdot k_T)^3 \bigg( k_T^2 \big( ((7-3x)x-2) l_T^2 + 3 M^2 x (x-1)^2 - (2x^3+x) M_S^2 \big) \notag \\
& + 2 (x-1) k_T^4 + x \big( M^2 (x-1)^2 - M_S^2 \big) \big( -(x-2) l_T^2 + M^2 (x-1)^2 + (-2(x-1)x-1) M_S^2 \big) \bigg) \notag \\
& + k_T^4 l_T^4 \bigg( k_T^2 \big( -(x-2) l_T^2 + 2 M^2 (x-1)^2 - (x^2-2) M_S^2 \big) + 2 k_T^4 \notag \\
& - x M_S^2 \big( (2x-3) l_T^2 + M^2 (x-2) (x-1)^2 + (2x^2+x-2) M_S^2 \big) \bigg) \notag \\
& - 4 M_S^2 (l_T \cdot k_T)^4 \big( (x-1) k_T^2 + M^2 x (x-1)^2 - x M_S^2 \big) \notag \\
& + k_T^2 l_T^2 (l_T \cdot k_T)^2 \bigg( k_T^2 \big( -(x-3) l_T^2 + 2 M^2 (x-1)^2 + (x(5x-14)+6) M_S^2 \big) + k_T^4 \notag \\
& + M_S^2 \big( 2x(2x-3) l_T^2 + M^2 (x(3x-4)-2) (x-1)^2 \big) + M^4 (x-1)^4 + \big( x(4x^2+x-4)+1 \big) M_S^4 \bigg) \notag \\
& + k_T^2 l_T^2 l_T \cdot k_T \bigg( k_T^4 \big( -(x-5) l_T^2 + 2 M^2 (x-1)^2 + (x-2)(2x-1) M_S^2 \big) \notag \\
& + k_T^2 \big( 2 l_T^2 \big( M^2 (x-1)^2 + (-x^2+x+1) M_S^2 \big) - (x-2) l_T^4 + M^4 (x-1)^4 \notag \\
& + 2 M^2 (x+1) (x-1)^3 M_S^2 + (1-2x^3) M_S^4 \big) + k_T^6 \notag \\
& + x M_S^2 \big( M^2 (x-1)^2 - M_S^2 \big) \big( -(x-2) l_T^2 + M^2 (x-1)^2 + (-2(x-1)x-1) M_S^2 \big) \bigg) \bigg\}\left(16 \pi^3 x k_T^6 M_S^4\right)^{-1}\,,
\end{align}

\begin{align}
&\mathcal{C}_{221}^{[h_{1T}^\perp]}+\mathcal{C}_{212}^{[h_{1T}^\perp]}+\mathcal{C}_{122}^{[h_{1T}^\perp]}\notag\\
=& 3 M^2 (x-1) \Bigg\{ k_T^4 l_T^2 \bigg( k_T^2 \big( 4 l_T^2 + (x+4) M_S^2 \big)  + x M_S^2 \big( 3 l_T^2 + M^2 (x-1)^2 - M_S^2 \big) \bigg) \notag \\
& + k_T^2 (l_T \cdot k_T)^2 \bigg( -M_S^2 \big( (5x-8) k_T^2 + (6x+1) l_T^2 + 3 M^2 x (x-1)^2 \big)  + l_T^2 \big( 4 k_T^2 + M^2 (x-1)^2 \big) + 5 x M_S^4 \bigg) \notag \\
& + k_T^2 l_T \cdot k_T \bigg( k_T^2 \big( M_S^2 \big( 3 (x+1) l_T^2 - 2 M^2 (x-1)^2 x \big) + M^2 (x-1)^2 l_T^2 + 4 l_T^4 + 4 x M_S^4 \big) \notag \\
& + k_T^4 \big( 4 l_T^2 - 2 (x-3) M_S^2 \big) + 2 x l_T^2 M_S^2 \big( M^2 (x-1)^2 - M_S^2 \big) \bigg) \notag \\
& + 2 M_S^2 (l_T \cdot k_T)^3 \big( (1-4x) k_T^2 + 2 x (M_S^2 - M^2 (x-1)^2) \big) \Bigg\}\left(4 \pi^3 x k_T^6 M_S^4\right)^{-1}\,,
\end{align}

\begin{align}
&\mathcal{C}_{233}^{[h_{1T}^\perp]}+\mathcal{C}_{323}^{[h_{1T}^\perp]}+\mathcal{C}_{332}^{[h_{1T}^\perp]}\notag\\
=& 3 \bigg\{ -2 \big( k_T^2 + l_T \cdot k_T \big) \bigg( -l_T^2 M_S^4 - \big( l_T^2 + 2 l_T \cdot k_T \big)^2 M_S^2 \notag \\
& + \big( M^2 (x-1)^2 + k_T^2 \big) l_T^2 \big( M^2 (x-1)^2 + k_T^2 + l_T^2 + 2 l_T \cdot k_T \big) \bigg) \big( k_T^2 l_T^2 + l_T \cdot k_T (M^2 (x-1)^2 + k_T^2 - M_S^2) \big) k_T^2 \notag \\
& + (x-1) \bigg[ k_T^4 \big( M^2 (x-1)^2 + k_T^2 - M_S^2 \big) \big( M^2 (x-1)^2 + k_T^2 + 2 l_T^2 + 3 M_S^2 \big) l_T^4 \notag \\
& + 2 l_T \cdot k_T \, k_T^2 \Big( -\big( (8 k_T^2 + (x+1) l_T^2) M_S^4 \big) + l_T^2 \big( M^2 x (x-1)^2 + (x-5) k_T^2 - l_T^2 \big) M_S^2 \notag \\
& + \big( M^2 (x-1)^2 + k_T^2 \big) l_T^2 \big( M^2 (x-1)^2 + 2 k_T^2 + l_T^2 \big) \Big) l_T^2 \notag \\
& + 4 (l_T \cdot k_T)^3 M_S^2 \Big( 2 x M_S^4 + \big( -2 M^2 x (x-1)^2 - 2 (x+2) k_T^2 + l_T^2 \big) M_S^2  - \big( M^2 (x-1)^2 + 3 k_T^2 \big) l_T^2 \Big) \notag \\
& + (l_T \cdot k_T)^2 \Big( \big( 3 k_T^4 + 4 \big( M^2 (x-1)^2 - 3 M_S^2 \big) k_T^2 + \big( M_S^2 - M^2 (x-1)^2 \big)^2 \big) l_T^4 \notag \\
& + 2 M_S^2 \big( (x-6) k_T^4 + 2 \big( M^2 (x-1)^3 - 3 (x+1) M_S^2 \big) k_T^2 + x \big( M_S^2 - M^2 (x-1)^2 \big)^2 \big) l_T^2  - 16 k_T^4 M_S^4 \Big) \bigg] k_T^2 \notag \\
& - (x-1) \bigg[ \big( k_T^2 + l_T \cdot k_T \big) l_T^2 \big( M^2 (x-1)^2 + k_T^2 - l_T^2 \big) k_T^4  - 2 x l_T \cdot k_T \big( l_T \cdot k_T (k_T^2 + 2 l_T \cdot k_T) - k_T^2 l_T^2 \big) M_S^4 \notag \\
& + \Big( -l_T^2 \big( k_T^2 + 2 x l_T^2 \big) k_T^4 + 2 (l_T \cdot k_T)^2 \big( M^2 x (x-1)^2 + (x-4) k_T^2 + 2 x l_T^2 \big) k_T^2 \notag \\
& + l_T \cdot k_T \big( -4 k_T^4 - \big( 2 M^2 x (x-1)^2 + k_T^2 \big) l_T^2 \big) k_T^2 \notag \\
& + 4 (x-1) (l_T \cdot k_T)^3 \big( (x-1) x M^2 + k_T^2 \big) \Big) M_S^2 \bigg] \Big( l_T^4 + \big( M^2 (x-1)^2 + k_T^2 + 3 M_S^2 + 2 l_T \cdot k_T \big) l_T^2 \notag \\
& + 4 l_T \cdot k_T \, M_S^2 \Big) \bigg\}\left(64 M^2 \pi^3 (x-1) x k_T^6 M_S^4\right)^{-1}\,,
\end{align}

\begin{align}
&\mathcal{C}_{123}^{[h_{1T}^\perp]}+\mathcal{C}_{132}^{[h_{1T}^\perp]}+\mathcal{C}_{213}^{[h_{1T}^\perp]}+\mathcal{C}_{231}^{[h_{1T}^\perp]}+\mathcal{C}_{321}^{[h_{1T}^\perp]}+\mathcal{C}_{312}^{[h_{1T}^\perp]}\notag\\
=& 3 \Bigg\{ 6 l_T^4 k_T^8 + l_T^2 \Big( -3 (x-2) l_T^4 + 6 \big( M^2 (x-1)^2 - (x-2) M_S^2 \big) l_T^2 \notag \\
& - 8 (x-1) M_S^4 \Big) k_T^6 + 2 x l_T^4 M_S^2 \Big( M^2 (x-1)^2 - (x-2) l_T^2 + (3-2x) M_S^2 \Big) k_T^4 \notag \\
& + (l_T \cdot k_T)^2 \bigg( -\Big( \big( M^2 (x-3) (x-1)^2 + (3x-11) k_T^2 + ((5-4x)x+3) M_S^2 \big) l_T^4 \Big) \notag \\
& + \Big( 5 M^4 (x-1)^4 + 14 M^2 k_T^2 (x-1)^2 - 2 \big( (x^2-1) M^2 + 7 k_T^2 \big) M_S^2 (x-1) + 9 k_T^4 \notag \\
& + (8x^2-6x-3) M_S^4 \Big) l_T^2 + 2 M_S^2 \Big( M^4 (x-1)^4 + 4 M^2 k_T^2 (x-1)^2 + 3 k_T^4 + (4x+1) M_S^4 \notag \\
& - 2 \big( M^2 (2x+1) (x-1)^2 + 2 (x-4) k_T^2 \big) M_S^2 \Big) \bigg) k_T^2  - 8 (l_T \cdot k_T)^4 M_S^2 \Big( M^2 x (x-1)^2 + k_T^2 (x-1) - x M_S^2 \Big) \notag \\
& - 4 (l_T \cdot k_T)^3 \bigg( -\Big( \big( l_T^2 + (3-2x) M_S^2 \big) k_T^4 \Big) + \Big( (-2x^2+x-3) M_S^4 \notag \\
& + M^2 (x-1)^2 (3x-1) M_S^2 + l_T^2 \big( -M^2 (x-1)^2 - ((x-4)x+1) M_S^2 \big) \Big) k_T^2 \notag \\
& + x M_S^2 \big( M^2 (x-1)^2 - M_S^2 \big) \big( M^2 (x-1)^2 - (x-2) l_T^2 + (3-2x) M_S^2 \big) \bigg) \notag \\
& + l_T \cdot k_T \bigg( \big( 5 l_T^2 + 2 M_S^2 \big) k_T^8 + \Big( (17-3x) l_T^4 + 2 \big( 5 M^2 (x-1)^2 + (5-3x) M_S^2 \big) l_T^2 \notag \\
& + (20-8x) M_S^4 + 4 M^2 (x-1)^2 M_S^2 \Big) k_T^6 + \Big( -3 (x-2) l_T^6 \notag \\
& + \big( (-2x^2+x+9) M_S^2 - M^2 (x-9) (x-1)^2 \big) l_T^4 + \big( 5 M^4 (x-1)^4 + 2 M^2 (2x-1) M_S^2 (x-1)^2 \notag \\
& + (5-4(x-1)x) M_S^4 \big) l_T^2 + (8x+2) M_S^6 - 4 M^2 (x-1)^2 (2x+1) M_S^4 + 2 M^4 (x-1)^4 M_S^2 \Big) k_T^4 \notag \\
& + 2 x l_T^2 M_S^2 \big( M^2 (x-1)^2 - M_S^2 \big) \big( M^2 (x-1)^2 - (x-2) l_T^2 + (3-2x) M_S^2 \big) k_T^2 \bigg) \Bigg\}\left(16 \pi^3 x k_T^6 M_S^4\right)^{-1}\,,
\end{align}

\subsection{LT polarized helicity function \texorpdfstring{$g_{1LT}$}{g1LT}}
Since $g_{1LT}$ depends on the coupling constant $\kappa_2$, the number of nonvanishing terms for $g_{1LT}$ are significantly smaller than those of vector polarized gluon TMDs. 

\begin{align}
&\mathcal{C}_{222}^{[g_{1LT}]}\notag\\
=& 3 \Bigg\{ k_T^2 l_T^2 \Big( l_T^2 \big( M^2 (x-1)^2 - M_S^2 \big) - M_S^2 \big( k_T^2 - M^2 (x-1)^2 + M_S^2 \big) \Big) \notag \\
& + (l_T \cdot k_T)^2 \Big( 2 l_T^2 \big( k_T^2 + M^2 (x-1)^2 - 2 M_S^2 \big) + M_S^2 \big( (2x-3) k_T^2 + M^2 (x-1)^2 - M_S^2 \big) \Big) \notag \\
& + l_T \cdot k_T \bigg( k_T^2 \Big( l_T^2 \big( 3 M^2 (x-1)^2 + (x-3) M_S^2 \big) + l_T^4 + 2 M^2 (x-1)^2 M_S^2 - 2 M_S^4 \Big) \notag \\
& + k_T^4 \big( l_T^2 + (x-2) M_S^2 \big) - \big( M^2 (x-1)^2 - M_S^2 \big) \Big( M_S^2 \big( M^2 (x-1)^2 x - l_T^2 \big) - l_T^4 - x M_S^4 \Big) \bigg) \notag \\
& - 4 M_S^2 (l_T \cdot k_T)^3 \Bigg\}\left(16 \pi^3 k_T^2 M_S^4\right)^{-1}\,,
\end{align}

\begin{align}
&\mathcal{C}_{112}^{[g_{1LT}]}+\mathcal{C}_{211}^{[g_{1LT}]}\notag\\
=& 3 \Bigg\{ 2 (x-1) k_T^2 l_T^2 \Big( M^2 (x-1) l_T^2 + 2 M_S^2 \big( M^2 (x-1) + M_S^2 \big) \Big) \notag \\
& + 2 (l_T \cdot k_T)^2 \Big( l_T^2 \big( k_T^2 + M^2 (x-1)^2 - 3 M_S^2 \big) - 2 M_S^2 \big( k_T^2 - M^2 (x-1)^2 + M_S^2 \big) \Big) \notag \\
& + l_T \cdot k_T \bigg( l_T^4 \big( k_T^2 + M^2 (x-1)^2 - M_S^2 \big) + l_T^2 \Big( 4 M^2 (x-1)^2 k_T^2 + k_T^4 - \big( M_S^2 - M^2 (x-1)^2 \big)^2 \Big) \notag \\
& + 2 M_S^2 \Big( 2 M_S^2 \big( (x-2) k_T^2 + M^2 (x-1)^3 \big) + 4 M^2 (x-1)^2 k_T^2 - k_T^4 + M^4 (x-1)^4 + (1-2x) M_S^4 \Big) \bigg) \notag \\
& - 8 M_S^2 (l_T \cdot k_T)^3 \Bigg\}\left(16 \pi^3 k_T^2 M_S^4\right)^{-1}\,,
\end{align}

\begin{align}
&\mathcal{C}_{332}^{[g_{1LT}]}+\mathcal{C}_{233}^{[g_{1LT}]}\notag\\
=& 3 \Bigg\{ \Big( l_T^2 \big( k_T^2 + M^2 (x-1)^2 - M_S^2 \big) - 4 M_S^2 l_T \cdot k_T \Big)  \Big( l_T \cdot k_T \big( 3 k_T^2 + l_T^2 - M^2 (x-1)^2 + M_S^2 \big) + k_T^2 l_T^2 + 2 (l_T \cdot k_T)^2 \Big) \notag \\
& \times \Big( l_T^2 \big( 2 l_T \cdot k_T + k_T^2 + M^2 (x-1)^2 + 3 M_S^2 \big) + 4 M_S^2 l_T \cdot k_T + l_T^4 \Big) \Bigg\}\left(256 \pi^3 M^4 (1-x) k_T^2 M_S^4\right)^{-1}\,,\label{gLT:233}
\end{align}

\begin{align}
&\mathcal{C}_{122}^{[g_{1LT}]}+\mathcal{C}_{212}^{[g_{1LT}]}+\mathcal{C}_{221}^{[g_{1LT}]}\notag\\
=& 3 \Bigg\{ k_T^2 l_T^2 \bigg( 2 M_S^2 \big( k_T^2 - 2 M^2 (x-1)^2 + x M_S^2 \big)  + l_T^2 \big( M_S^2 - 3 M^2 (x-1)^2 \big) \bigg) \notag \\
& + 2 (l_T \cdot k_T)^2 \bigg( -M_S^2 \big( (2x-5) k_T^2 - 5 l_T^2 + M^2 (x-1)^2 \big)  - 2 l_T^2 \big( k_T^2 + M^2 (x-1)^2 \big) + M_S^4 \bigg)\notag\\
&  + l_T \cdot k_T \bigg( -2 l_T^4 \big( k_T^2 + M^2 (x-1)^2 - M_S^2 \big)  + l_T^2 \Big( k_T^2 \big( (5-2x) M_S^2 - 7 M^2 (x-1)^2 \big) - 2 k_T^4 + \big( M_S^2 - M^2 (x-1)^2 \big)^2 \Big) \notag \\
& + 2 M_S^2 \Big( k_T^2 \big( (x+1) M_S^2 - 3 M^2 (x-1)^2 \big) - (x-3) k_T^4  + M^4 x (x-1)^4 - M^2 x (x-1)^2 M_S^2 \Big) \bigg) \notag \\
& + 12 M_S^2 (l_T \cdot k_T)^3 \Bigg\}\left(16 \pi^3 k_T^2 M_S^4\right)^{-1}\,,
\end{align}

\begin{align}
&\mathcal{C}_{322}^{[g_{1LT}]}+\mathcal{C}_{223}^{[g_{1LT}]}\notag\\
=& 3 \Bigg\{ \Big[ l_T^2 \big( k_T^2 + M^2 (x-1)^2 \big) \big( 2 l_T \cdot k_T + k_T^2 + l_T^2 + M^2 (x-1)^2 \big)  - M_S^2 \big( 2 l_T \cdot k_T + l_T^2 \big)^2 - l_T^2 M_S^4 \Big] \notag\\
&\times \Big[ l_T \cdot k_T \big( 3 k_T^2 + l_T^2 - M^2 (x-1)^2 + M_S^2 \big)  + k_T^2 l_T^2 + 2 (l_T \cdot k_T)^2 \Big] \Bigg\}\left(64 \pi^3 M^2 (x-1) k_T^2 M_S^4\right)^{-1}\,,\label{gLT:223}
\end{align}

\begin{align}
&\mathcal{C}_{232}^{[g_{1LT}]}\notag\\
=& -3 \Bigg\{ k_T^2 l_T^4 \bigg( l_T^2 \big( M^2 (x-1)^2 - M_S^2 \big)  - 2 M_S^2 \big( k_T^2 - M^2 (x-1)^2 + M_S^2 \big) \bigg) \notag \\
& + 2 l_T^2 (l_T \cdot k_T)^2 \bigg( l_T^2 \big( k_T^2 + M^2 (x-1)^2 - 3 M_S^2 \big)  + M_S^2 \big( (2x-3) k_T^2 + M^2 (x-1)^2 - (2x+5) M_S^2 \big) \bigg) \notag \\
& + l_T^2 l_T \cdot k_T \bigg( k_T^2 \Big( l_T^2 \big( 3 M^2 (x-1)^2 + (2x-3) M_S^2 \big) + l_T^4  + 4 (x-1) M_S^2 \big( M^2 (x-1) + M_S^2 \big) \Big)  \notag \\
&+ k_T^4 \big( l_T^2 + 2 (x-2) M_S^2 \big) - \big( M^2 (x-1)^2 - M_S^2 \big) \Big( 2 M_S^2 \big( M^2 (x-1)^2 x - l_T^2 \big) - l_T^4 + 2 x M_S^4 \Big) \bigg) \notag \\
& - 8 M_S^2 (l_T \cdot k_T)^3 \big( l_T^2 + (x+1) M_S^2 \big) \Bigg\}\left(64 \pi^3 M^2 k_T^2 M_S^4\right)^{-1}\,,
\end{align}

\begin{align}
&\mathcal{C}_{213}^{[g_{1LT}]}+\mathcal{C}_{312}^{[g_{1LT}]}\notag\\
=& 3\Bigg\{ M_S^4 \Big( -4 l_T \cdot k_T - 4 k_T^2 - l_T^2 + 4 M^2 (x-1)^2 \Big)  - M_S^2 \bigg( l_T \cdot k_T \Big( 4 k_T^2 + 6 l_T^2 + 4 M^2 (x-1)^2 \Big) + 8 (l_T \cdot k_T)^2 \notag \\
& + 2 \big( k_T^2 + M^2 (x-1)^2 \big)^2 + l_T^4 \bigg)  + l_T^2 \big( k_T^2 + M^2 (x-1)^2 \big) \Big( 2 l_T \cdot k_T + k_T^2 + l_T^2 + M^2 (x-1)^2 \Big)- 2 M_S^6 \Bigg\} \notag \\
&  \times \Bigg\{ l_T \cdot k_T \Big( 3 k_T^2 + l_T^2 - M^2 (x-1)^2 + M_S^2 \Big)  + k_T^2 l_T^2 + 2 (l_T \cdot k_T)^2 \Bigg\}\left(64 \pi^3 M^2 (1-x) k_T^2 M_S^4\right)^{-1}\,,\label{gLT:213}
\end{align}

\begin{align}
&\mathcal{C}_{132}^{[g_{1LT}]}+\mathcal{C}_{231}^{[g_{1LT}]}\notag\\ 
=&  \Bigg\{ 6 M^2 (x-1)^2 k_T^2 l_T^4 \big( l_T^2 + 4 M_S^2 \big) \notag \\
& + 6 l_T^2 (l_T \cdot k_T)^2 \Big( l_T^2 \big( k_T^2 + M^2 (x-1)^2 - 3 M_S^2 \big) + 4 M^2 (x-1)^2 M_S^2 - 8 M_S^4 \Big) \notag \\
& + 3 l_T^2 l_T \cdot k_T \bigg( l_T^4 \big( k_T^2 + M^2 (x-1)^2 - M_S^2 \big) \notag \\
&  + l_T^2 \Big( 2 M_S^2 \big( k_T^2 + 2 M^2 (x-1)^2 \big) + 4 M^2 (x-1)^2 k_T^2 + k_T^4 - M^4 (x-1)^4 - 3 M_S^4 \Big) \notag \\
&  + 16 M^2 (x-1)^2 k_T^2 M_S^2 \bigg)  - 24 M_S^2 (l_T \cdot k_T)^3 \big( l_T^2 + 2 M_S^2 \big) \Bigg\}\left(64 \pi^3 M^2 k_T^2 M_S^4\right)^{-1}\,.
\end{align}

\subsection{TT polarized worm-gear function \texorpdfstring{$g_{1TT}$}{g1TT}}
When $ijk=\{312,322,332,213,223,233\}$, we can obtain $\mathcal{C}_{ijk}^{[g_{1TT}]}$ by using $\mathcal{C}_{ijk}^{[g_{1LT}]}$ in Eqs.~\eqref{gLT:233},~\eqref{gLT:223},~\eqref{gLT:213}:
\begin{align}
\mathcal{C}_{\{312,322,332\}}^{[g_{1TT}]}=&\frac{2M^2 (x-1)}{k_T^2 + M_S^2 -M^2 (1-x)^2}\mathcal{C}_{\{312,322,332\}}^{[g_{1LT}]}\,,\\
\mathcal{C}_{\{213,223,233\}}^{[g_{1TT}]}=&\frac{2(k_T^2+l_T \cdot k_T)M^2(x-1)}{k_T^2 ((k_T+l_T)^2-M^2(1-x)^2+M_S^2)}\mathcal{C}_{\{213,223,233\}}^{[g_{1LT}]}\,.
\end{align}
Then,
\begin{align}
&\mathcal{C}_{222}^{[g_{1TT}]}\notag\\
=& 3 M^2 (x-1) \Bigg\{ k_T^2 l_T \cdot k_T \bigg( M_S^2 \big( 2 (x-2) k_T^2 + (x-2) l_T^2 + 2 M^2 x (x-1)^2 \big)  + 2 M^2 (x-1)^2 l_T^2 - 2 x M_S^4 \bigg)\notag\\
& + (l_T \cdot k_T)^2 \bigg( l_T^2 \big( k_T^2 + M^2 (x-1)^2 - M_S^2 \big)  + M_S^2 \big( (3x-4) k_T^2 + M^2 x (x-1)^2 - x M_S^2 \big) \bigg)\notag\\
& - k_T^4 l_T^2 \big( l_T^2 + 2 M_S^2 \big)  - 2 M_S^2 (l_T \cdot k_T)^3 \Bigg\}\left(16 \pi^3 k_T^4 M_S^4\right)^{-1}\,,
\end{align}

\begin{align}
&\mathcal{C}_{112}^{[g_{1TT}]}+\mathcal{C}_{211}^{[g_{1TT}]}\notag\\
=& 3 M^2 (x-1) \Bigg\{ l_T^2 l_T \cdot k_T \Big( l_T \cdot k_T \big( k_T^2 + M^2 (x-1)^2 \big) + 2 M^2 (x-1)^2 k_T^2 \Big) \notag \\
& - M_S^2 \bigg( 2 k_T^4 l_T^2 + 4 k_T^4 l_T \cdot k_T + 4 (l_T \cdot k_T)^3 + \big( 4 k_T^2 + l_T^2 \big) (l_T \cdot k_T)^2 \bigg)  - k_T^4 l_T^4 \Bigg\}\left(8 \pi^3 k_T^4 M_S^4\right)\,,
\end{align}

\begin{align}
&\mathcal{C}_{122}^{[g_{1TT}]}+\mathcal{C}_{212}^{[g_{1TT}]}+\mathcal{C}_{221}^{[g_{1TT}]}\notag\\   
=& 3 M^2 (1-x) \Bigg\{ 2 k_T^2 l_T \cdot k_T \bigg( M_S^2 \big( 2 (x-3) k_T^2 + (x-2) l_T^2 + 2 M^2 x (x-1)^2 \big)  + 3 M^2 (x-1)^2 l_T^2 - 2 x M_S^4 \bigg) \notag \\
& + (l_T \cdot k_T)^2 \bigg( 3 l_T^2 \big( k_T^2 + M^2 (x-1)^2 - M_S^2 \big) + 2 M_S^2 \big( 3 (x-2) k_T^2 + M^2 x (x-1)^2 - x M_S^2 \big) \bigg) \notag \\
& - 3 k_T^4 l_T^2 \big( l_T^2 + 2 M_S^2 \big)  - 8 M_S^2 (l_T \cdot k_T)^3 \Bigg\}\left(16 \pi^3 k_T^4 M_S^4\right)^{-1}\,,
\end{align}

\begin{align}
&\mathcal{C}_{232}^{[g_{1TT}]}\notag\\
=& 3 (1-x) \Bigg\{ k_T^2 l_T^2 l_T \cdot k_T \bigg( l_T^2 \Big( l_T \cdot k_T + 2 (x-1) \big( M^2 (x-1) + M_S^2 \big) \Big) \notag \\
& + 2 M_S^2 \Big( (3x-4) l_T \cdot k_T + 2 x \big( M^2 (x-1)^2 + M_S^2 \big) \Big) \bigg) \notag \\
& + (l_T \cdot k_T)^2 \big( l_T^2 + 2 x M_S^2 \big) \Big( M^2 (x-1)^2 l_T^2 - M_S^2 \big( 4 l_T \cdot k_T + l_T^2 \big) \Big) \notag \\
& - k_T^4 \Big( -4 (x-2) l_T^2 M_S^2 l_T \cdot k_T + 4 l_T^4 M_S^2 + l_T^6 \Big) \Bigg\}\left(64 \pi^3 k_T^4 M_S^4\right)^{-1}\,,
\end{align}

\begin{align}
&\mathcal{C}_{132}^{[g_{1TT}]}+\mathcal{C}_{231}^{[g_{1TT}]}\notag\\ 
=& 3(x-1) \Bigg\{ 2 k_T^2 l_T^2 l_T \cdot k_T \Big( 2 M_S^2 \big( -k_T^2 + M^2 (x-1)^2 + M_S^2 \big) + M^2 (x-1)^2 l_T^2 \Big) \notag \\
& + l_T^2 (l_T \cdot k_T)^2 \bigg( l_T^2 \big( k_T^2 + M^2 (x-1)^2 - M_S^2 \big) - 2 M_S^2 \big( k_T^2 - M^2 (x-1)^2 + M_S^2 \big) \bigg) \notag \\
& - k_T^4 l_T^4 \big( l_T^2 + 4 M_S^2 \big)  - 4 M_S^2 (l_T \cdot k_T)^3 \big( l_T^2 + 2 M_S^2 \big) \Bigg\}\left(32\pi^3 k_T^4 M_S^4\right)^{-1}\,.
\end{align}

\section{Positivity bounds}\label{appendix2}
At leading twist, the positivity bounds are the following nine inequalities:
\begin{align}
\frac{\bm{k}_T^2}{2M^2}\left|h_1^\perp-h_{1LL}^\perp\right|&\leq f_1-f_{1LL}\,,\\
\frac{\bm{k}_T^4}{16M^4}\left[4\left(h_{1L}^\perp\right)^2+\left(2h_{1}^\perp+h_{1LL}^\perp\right)^2\right]&\leq \left(f_1+\frac{f_{1LL}}{2}+g_1\right)\left(f_1+\frac{f_{1LL}}{2}-g_1\right)\,,\\
\frac{\bm{k}_T^2}{2M^2}\left(h_1^2+4h_{1LT}^2\right)&\leq \left(f_1-f_{1LL}\right) \left(f_1+\frac{f_{1LL}}{2}+g_1\right)\,,\\
\frac{\bm{k}_T^6}{8M^6}\left[\left(h_{1T}^\perp\right)^2+\left(h_{1LT}^\perp\right)^2\right]&\leq \left(f_1-f_{1LL}\right) \left(f_1+\frac{f_{1LL}}{2}-g_1\right)\,,\label{eq:bound2}\\
\frac{\bm{k}_T^2}{2M^2}\left[\left(f_{1T}^\perp +g_{1LT}\right)^2+\left(f_{1LT}+g_{1T}+h_{1LT}\right)^2\right]&\leq \left(f_1-f_{1LL}\right) \left(f_1+\frac{f_{1LL}}{2}+g_1\right)\,,\label{eq:bound1}\\
\frac{\bm{k}_T^2}{2M^2}\left[\left(f_{1T}^\perp -g_{1LT}\right)^2+\left(f_{1LT}-g_{1T}+h_{1LT}\right)^2\right]&\leq \left(f_1-f_{1LL}\right) \left(f_1+\frac{f_{1LL}}{2}-g_1\right)\,,\\
\left|h_{1TT}\right|&\leq \frac{1}{2}\left(f_1+\frac{f_{1LL}}{2}+g_1\right)\,,\\
\frac{\bm{k}_T^4}{2M^4}\left|h_{1TT}^{\perp \perp}\right|&\leq \left(f_1+\frac{f_{1LL}}{2}-g_1\right)\,,\\
\frac{\bm{k}_T^4}{M^4}\left[g_{1TT}^2+\left(f_{1TT}-h_{1TT}^\perp\right)^2\right]&\leq \left(f_1+\frac{f_{1LL}}{2}+g_1\right)\left(f_1+\frac{f_{1LL}}{2}-g_1\right)\,.
\end{align}

For integrated TMDs case, there are the following three bounds:
\begin{align}
\left|g_1\right|&\leq f_1 +\frac{f_{1LL}}{2}\,,\\
f_{1LL}&\leq f_1\,,\\
\left|h_{1TT}\right|&\leq \frac{1}{2} \left(f_1+\frac{f_{1LL}}{2}+g_1\right)\,.
\end{align}

\end{widetext}

\end{document}